\documentclass[9pt,a4paper,twocolumn]{IEEEtran}

\usepackage{graphicx}
\usepackage[comma,numbers,square,sort&compress]{natbib}
\usepackage{epstopdf}
\usepackage{amsmath}
\usepackage{enumerate}
\usepackage{subfigure}
\usepackage{ntheorem}
\usepackage{amssymb}
\usepackage{mathrsfs}        
\usepackage{indentfirst}

\newtheorem{definition}{Definition}
\newtheorem{lemma}{Lemma}
\newtheorem{theorem}{Theorem}
\newtheorem{corollary}{Corollary}

\begin{document}

\title{Frequency Domain Identifiability and Sloppiness of Descriptor Systems with an LFT Structure} 

\author{Tong Zhou 
\thanks{This work was supported in part by the NNSFC under Grant 61733008,  52061635102 and 61573209.}
\thanks{Tong Zhou is with the Department of Automation, Tsinghua University, Beijing, 100084, P.~R.~China
        {(email: {\tt\small tzhou@mail.tsinghua.edu.cn}.)}%
}}

\maketitle

\begin{abstract}       
Identifiability and sloppiness are investigated in this paper for the parameters of a descriptor system based on its frequency response samples. Two metrics are suggested respectively for measuring absolute and relative sloppiness of the parameter vector at a prescribed value. In this descriptor system, system matrices are assumed to depend on its parameters through a linear fractional transformation (LFT). When an associated transfer function matrix (TFM) is of full normal row rank, a matrix rank based necessary and sufficient condition is derived for parameter identifiability with a set of finitely many frequency responses. This condition can be verified recursively which is computationally quite appealing, especially when the system is of a large scale. From this condition, an algorithm is suggested to find a set of frequencies with which the frequency responses of the system are capable to uniquely determine its parameters. An ellipsoid approximation is given for the set consisting of all the parameter values with which the associated descriptor system has a frequency response that deviates within a prescribed distance, from that corresponding to a globally identifiable parameter vector value. Explicit formulas are also derived for the suggested absolute and relative sloppiness metrics.
\end{abstract}

\renewcommand{\labelenumi}{\rm\bf A\arabic{enumi})}



\section{Introduction}

Model building from experiment data is fundamental in system analysis and synthesis, which has attracted extensive attentions for a long time in various fields including engineering, economy, biology, etc., and various approaches have been suggested and numerous results have been obtained \cite{Ljung1999,mmrt1992,ps2001,Verhaegen1996,vd1996,zyl2018}. When a dynamic model is to be estimated, the developed approaches can be briefly categorized into the time domain approach and the frequency domain approach, in which time domain input/output experiment data and their Fourier transformations or system frequency response samples are respectively utilized.

To estimate a parameter from experiment data, three aspects must be taken into account. That is, identifiability of the parameter, informativity of experiment data, and efficiency of an estimation algorithm \cite{adm2020, Ljung1999, cw2020,wvd2018,zy2022}. Briefly, identifiability requires existence of a probing signal which can differentiate outputs of the system with different parameter values, while a set of experiment data is called informative if it can uniquely determine the parameter value. Note that in actual applications, only noisy and finitely many experiment data can be collected. In addition, there are also situations in which a probing signal must meet some implementation constraints. It is sometimes called practical identifiability when a parameter can be recovered within a prescribed precision using a corrupted data set of a finite length under some restricted experiment conditions. It is argued in \cite{gjmabchjkrs2019} and \cite{whrtt2021} that practical identifiability is in general more challenging and has not been well investigated yet.

On the other hand, when there is a great number of parameters that are to be identified simultaneously from a set of experiment data, which is often encountered in large scale networked dynamic systems (NDS), it is observed in \cite{tmbdms2015}, \cite{wcgbmbes2006}, etc., that the trajectories of the system outputs corresponding to all possible parameter values universally  have a so-called hyper-ribbon structure, and influences of different parameter combinations on system outputs may be significantly different. In addition, there usually exist some parameter combinations, which are called sloppy parameter combinations, whose variations have a very limited influences on system outputs, even when they are identifiable. In other words, variations of these parameter combinations can hardly be observed from actual experiment data, meaning that they can are not practically identifiable. These phenomena happen in an NDS ranging from a cell signaling system and a brain neural network, to radioactive decays, economy systems and engineering systems. Occurrences of this phenomenon are argued there to have close relations with the eigenvalue distributions of the Fisher information matrix.  While these observations are extensively considered helpful in understanding behaviors of a large scale NDS and investigating data-driven system analysis and synthesis, there is still not a metric that can quantitatively measure sloppiness of a system parameter.

Descriptor systems are extensively utilized in system analysis and synthesis, especially when a system is constituted from a large number of subsystems \cite{Dai1989,Duan2010,Siljak1978,Zhou2020-2,zy2022}. One of their distinguished attractive properties lies in its capabilities of describing restrictions among different natural variables, which exist extensively in large scale NDSs. While analysis and synthesis of a descriptor system have been extensively studied, only a few works deal with its parameter identification. In particular, in \cite{mmrt1992} and \cite{Verhaegen1996}, some subspace based methods are suggested for estimating system matrices of a descriptor system using time-domain input-output data, while some necessary and sufficient conditions are given in \cite{zy2022} for its time-domain parameter identifiability.

In this paper, we investigate whether or not the parameters of a descriptor system can be uniquely determined by its frequency response at finitely many frequencies, and how many frequency response samples are required when the answer is positive, as well as how to determine the desirable frequencies. In addition, we also study how to quantify its sloppiness at a prescribed parameter vector value with these frequency response samples. In the adopted descriptor system, system matrices depend on its parameters through a linear fractional transformation (LFT). This model is argued in (\cite{Zhou2020-2}) and (\cite{zy2022}) to be able to represent uniformly influences of subsystem parameters and subsystem interactions on the input-output relations of an NDS, and therefore enables a uniform investigation on their estimations. Under the condition that the transfer function matrix (TFM) from system input vector to an auxiliary output vector is of full normal row rank (FNRR), a necessary and sufficient condition is derived for a finite set of frequencies with which the associated system frequency responses are able to uniquely determine the system parameters. This condition can be verified recursively, and is therefore computationally attractive. From this condition, an algorithm is suggested to search the desirable frequencies, which usually leads to a number of frequencies significantly smaller than that straightforwardly obtained from system orders. In addition, relations are also studied between variations of system frequency response and system parameters. Two metrics are suggested respectively for measuring absolute and relative sloppiness of a prescribed parameter vector value of the descriptor system. An ellipsoid approximation is explicitly derived for the set constituted from all the parameter vector values, with which the associated descriptor system has a frequency response deviating from that corresponding to a globally identifiable parameter vector value within a prescribed distance. From this approximation, analytic expressions are derived for an the suggested absolute and relative sloppiness metrics.

The following notation and symbols are adopted in this paper. $\mathcal{R}^m$ and $\mathcal{C}^{m\times n}$ stand respectively the set of $m\times n$ dimensional real and complex matrices. When $m=1$ and $n=1$, they are usually omitted, and the corresponding matrices reduced to row/column vectors or scalars. $\star_{r}^\bot$/$\star_{l}^\bot$ represents the matrix whose columns/rows form a base of the right/left null space of a matrix, while ${\rm\bf span}\{x_i|^n_{i=1}\}$ the space constituted from linear combinations of the vectors $x_i|^n_{i=1}$. ${\rm\bf diag}\{X_i|^n_{i=1}\}$ stands for a diagonal matrix with its $i$-th diagonal block being $X_i$, while ${\rm\bf col}\{X_i|^n_{i=1}\}$ the vector/matrix stacked by $X_i|^n_{i=1}$ with its $i$-th row block vector/matrix being $X_i$, and ${\rm\bf vec}\{X\}$ the vector stacked by the columns of matrix $X$. $\overline{\sigma}(\star)$ denotes the maximum singular value of a matrix, and ${\rm\bf det}(\star)$ the determinant of a square matrix, while $\star^{\dag}$ the Moore-Penrose inverse of a matrix. $||\star||_{2}$ stands for both the Euclidean norm of a column vector and its induced norm of a matrix, while $||\star||_{F}$ the Frobenius norm of a matrix. The superscript $T$ and $H$ represent respectively the transpose and the conjugate transpose of a matrix/vector. $I_{m}$ and $0_{m\times n}$ stands for the $m$ dimensional identity matrix and the $m\times n$ dimensional zero matrix. The subscripts are often omitted when this piece of information is not very essential.

\section{Problem Formulation and Preliminaries}\label{section:pfp}

Let $\delta$ represent either a differentiate operator with respect to time or a forward time shift operator. Concerning a lumped linear time invariant (LTI) descriptor system $\rm\bf\Sigma$, assume that its input-output relations are described by the following equations,
\begin{equation}\label{DesSys}
	E\delta(x(t))=A(\theta)x(t)+B(\theta)u(t),\quad y(t)=C(\theta)x(t)+D(\theta)u(t)
\end{equation}
with $\theta$ denoting the vector constituted from system parameters to be estimated. In addition, assume that its system matrices $A(\theta)$, $B(\theta)$, $C(\theta)$ and $D(\theta)$ depend on its parameters through a linear fractional transformation (LFT), which is expressed as follows,
\begin{eqnarray}\label{SysMat}
& & \hspace*{-1.0cm}	\left[\!\!\begin{array}{cc}
		A(\theta) & B(\theta) \\ C(\theta) & D(\theta)
	\end{array}\!\!\right] =
	\left[\!\!\begin{array}{cc}
		A_{xx} & B_{xu} \\
		C_{yx} & D_{yu}
	\end{array}\!\!\right] +
	\left[\!\!\begin{array}{c}
		B_{xv} \\ D_{yv}
	\end{array}\!\!\right]
    [I_{m_v}-P(\theta) D_{zv}]^{-1} \times \nonumber \\
& & \hspace*{4cm} P(\theta)
	\left[\!\!\begin{array}{cc}
		C_{zx} & D_{zu}
	\end{array}\!\!\right]
\end{eqnarray}
in which
\begin{equation}\label{eqn:ParVec}
	P(\theta) = \sum^{q}_{k=1}\theta_{k} P_{k} \hspace{0.5cm}{\rm with} \hspace{0.5cm} \theta = col\left\{ \theta_{k}|_{k=1}^{q} \right\} \in {\rm\bf\Theta}
\end{equation}

Throughout this paper, the numbers of rows and columns of the matrix $\star_{\#\S}$ with $\star$ being $A,B,C$ or $D$, and $\#,\S$ being $x$, $u$, $y$, $z$ or $v$, are denoted respectively by   $m_{\#}$ and $m_{\S}$. On the other hand, the TFM from the external input $u(t)$ of the descriptor system  $\Sigma$ to its external output $y(t)$ is denoted by $H(\lambda)$, while its frequency response by $H(j\omega)$, no matter it is continuous time or discrete time. This means that when $\delta$ is the time differentiate operator, $H(j\omega)$ is understood to be the value of the TFM $H(\lambda)$ with $\lambda = j\omega$ and $\omega \in {\cal R}$; while when $\delta$ is the time shift  operator, $H(j\omega)$ is understood to be the value of the TFM $H(\lambda)$ with $\lambda = e^{j\omega}$ and $\omega \in \left( -\pi,\; \pi \right]$.

In the above system dynamics descriptions, except the vector $\theta$, all the other matrices, that is, $E$, $A_{xx}$, $C_{yx}$, $B_{xu}$, $D_{yu}$, $B_{xv}$, $D_{yv}$, $D_{zv}$, $C_{zx}$, $D_{zu}$ and $P_{k}|_{k=1}^{q}$ are known and have a compatible dimension. These matrices are adopted to represent available knowledge about system dynamics and component connections, etc. Moreover, the set ${\rm\bf\Theta}$ reflects some available a priori information about the system parameters $\theta_{k}$, $k=1, 2, \cdots, q$. Some well adopted descriptions are that $\sum_{k=1}^{q}(\theta_{k}-\theta_{k,0})^{2} < \gamma$; or $|\theta_{k}-\theta_{k,0}| < \gamma$ for each $k=1, 2, \cdots, q$; in which $\theta_{k,0}|_{k=1}^{q}$ are known numbers, while $\gamma$ is a prescribed positive number, representing some available knowledge about system parameters  \cite{Ljung1999, zy2022, zyl2018}. Throughout this paper, however, it is assumed that $\theta_{k,0}=0$ for each $k=1,2,\cdots,q$. This is only for a concise presentation, and does not affect validness of the obtained results. As a matter of fact, using properties of LFTs, it is not very difficult to represent the nonzero $\theta_{k,0}$ with $k=1,2 \cdots, q$, though the system matrices $A_{xx}$, $B_{xu}$, $C_{yx}$ and $D_{yu}$.

In order to clarify dependence of the descriptor system $\rm\bf\Sigma$ on its parameter vector, it is sometimes also written as ${\rm\bf\Sigma}(\theta)$. Similar expressions are also adopted for its output vector $y(t)$, TFM $H(\lambda)$ and frequency response $H(j\omega)$, etc. In addition, in order to avoid awkward expressions, distinctions are not taken for the parameter vector $\theta$ and its value.

In this paper, the following 3 issues are dealt with for the descriptor system $\rm\bf\Sigma$.

\begin{itemize}
\item Possibilities to uniquely determine the value of its parameter vector $\theta$ from its frequency responses at finitely many frequencies. Moreover, determine the number of the required frequencies and the value of the desirable frequencies.
\item Computationally verifiable conditions for its parameter identifiability using a finite set of its frequency responses.
\item Quantify difficulties to identify the parameter vector $\theta$ from a finite set of its frequency responses.
\end{itemize}

Throughout this paper, the following two assumptions are adopted, which are respectively called the regularity assumption and the well-posedness assumption for brevity in the remaining of this paper.
\begin{enumerate}
\item The parameter set ${\rm\bf\Theta}$ is open, convex and contains $0$ as its element. In addition, the descriptor system of Eq.(\ref{DesSys}) is regular for each $\theta \in {\rm\bf\Theta}$, meaning that the matrix $E$ is square and ${\rm\bf det}\left\{ \lambda E - A(\theta) \right\} \not\equiv 0$.
\item The descriptor system $\rm\bf\Sigma$ is well-posed, meaning that the matrix $I_{m_v}- P(\theta) D_{zv}$ is invertible for every $\theta \in {\rm\bf\Theta}$.
\end{enumerate}

Briefly, these two assumptions means that system states and outputs respond in a deterministic way to an admissible external stimulus and an admissible initial state, and is essential for a descriptor system to work properly \cite{Dai1989,Duan2010,Kailath1980,Siljak1978,zdg1996,zyl2018}. Clearly, these requirements are also some prerequisites for estimating its parameters, implying that they are not very restrictive in actual applications. On the other hand, some computationally appealing necessary and sufficient conditions have been obtained in \cite{Zhou2020-2} for a large scale descriptor system to be regular.

To reveal the parameters $\theta_{k}$ with $k=1, 2, \cdots, q$, from experiment data, a prerequisite is that it must be identifiable. In addition, it is also essential to know what kind information, and how many data are required to reveal these parameters. Specifically, parameter identifiability can be defined as follows for the descriptor system $\rm\bf\Sigma$ of Eq.(\ref{DesSys}), which are consistent with those for parameter identifiability of a state space model, an NDS, etc., respectively adopted in \cite{Ljung1999}, \cite{ps2001},  \cite{zy2022} and \cite{Zhou2020-1}.

\begin{definition}\label{defi1}
$\;\;$
\begin{itemize}
\item Two distinct parameter vectors $\theta,\; \widetilde{\theta} \in {\rm\bf\Theta}$ are said to be differentiable, if for an arbitrary admissible initial state vector $x(0)$, there exists at least one external admissible input signal $u(t)|_{t=0}^{\infty}$, such that the external output $y(t,\theta)|_{t=0}^{\infty}$ of the descriptor system $\Sigma(\theta)$ is different from the external output $y(t,\widetilde{\theta})|_{t=0}^{\infty}$ of the descriptor system $\Sigma(\widetilde{\theta})$. Otherwise, these two parameter vectors are called undifferentiable.
\item The descriptor system $\rm\bf\Sigma$ is called globally identifiable at a specific parameter vector value $\theta$, if it is differentiable from any other parameter vector value $\widetilde{\theta} \in {\rm\bf\Theta}$. Otherwise, this parameter vector value $\theta$ is called globally unidentifiable.
\item The descriptor system $\rm\bf\Sigma$ is called locally identifiable at a specific parameter vector value $\theta$, if there exists an $\varepsilon$-neighborhood $\mathscr{B}(\theta,\varepsilon)$ which is a subset of ${\rm\bf\Theta}$, such that it is differentiable from any other parameter vector value $\widetilde{\theta}\in\mathscr{B}(\theta,\varepsilon)$. Otherwise, this parameter vector value $\theta$ is called locally unidentifiable.
\item The descriptor system $\rm\bf\Sigma$ is called locally/globally identifiable, if it is locally/globally identifiable at every $ \theta \in {\rm\bf\Theta}$.
\end{itemize}
\end{definition}

With these definitions, the following results can be directly obtained from  Theorem 1 in \cite{zy2022} and Theorem 1 in \cite{Zhou2020-1}.

\begin{lemma}\label{lemma:2}
Assume that the descriptor system $\rm\bf\Sigma$ satisfies Assumptions A1) and A2). Then it is globally identifiable at a specific parameter vector value  $\theta \in {\rm\bf\Theta}$, if and only if for every other element $\widetilde{\theta}$ of the set ${\rm\bf\Theta}$, the associated TFMs satisfy  $H(\lambda,\theta)\not\equiv  H(\lambda,\widetilde{\theta})$.
\end{lemma}

In addition to identifiability, it is also essential to investigate how difficult to estimate a parameter vector from experiment data, especially when its dimension is large, as there may exist sloppy parameter combinations whose variations almost do not change system input-output relations \cite{wcgbmbes2006,tmbdms2015}. For this purpose, the following concept is introduced in this paper, which quantitatively measure difficulties of identifying the parameters of the descriptor system  $\Sigma(\theta)$ from a finite set of its frequency response estimates.

\begin{definition}\label{defi2}
Let $\theta^{[0]}$ be an element of the set ${\rm\bf\Theta}$,  $\omega_{i}|_{i=1}^{N}$ some distinct frequencies, while $||\cdot||_{\star}$ a matrix norm and $||\cdot||_{\sharp}$ a vector norm. For a positive number $\varepsilon$, denote the following parameter set and its dimension respectively by $\widehat{\rm\bf\Theta}_{\star}(\varepsilon,\omega_{i}|_{i=1}^{N},\theta^{[0]})$ and $n_{\theta}$,
\begin{displaymath}
\left\{ \theta \;\left|\;\begin{array}{c}
\theta \in {\rm\bf\Theta} \\
\left|\left| {\rm\bf col}\left\{\left. H(j\omega_{i},\theta)\right|_{i=1}^{N}\right\} \!-\! {\rm\bf col}\left\{\left. H(j\omega_{i},\theta^{[0]})\right|_{i=1}^{N}\right\} \right|\right|_{\star}
 \!\leq\! \varepsilon \end{array}
\right.\!\right\}
\end{displaymath}
Moreover, define vectors $\theta_{v}^{[i]}|_{i=1}^{n_{\theta}}$ with the initial value $\theta_{v}^{[0]} = 0$ recursively as follows,
\begin{displaymath}
\theta_{v}^{[i]}
= \arg \max_{\theta\:\in \: \widehat{\rm\bf\Theta}_{\star}(\varepsilon,\omega_{i}|_{i=1}^{N},\theta^{[0]})\backslash{\rm\bf span} \left\{\theta_{v}^{[j]}|_{j=0}^{i-1} \right\}}      \left|\left| \theta - \theta^{[0]} \right|\right|_{\sharp}
\end{displaymath}
Then the following two quantities, denote them respectively by ${\rm\bf Sm}^{[a]}_{\star,\sharp}(\varepsilon,\omega_{i}|_{i=1}^{N},\theta^{[0]})$ and
${\rm\bf Sm}^{[r,k]}_{\star,,\sharp}(\varepsilon,\omega_{i}|_{i=1}^{N},\theta^{[0]})$ with $k=1,2,\cdots,n_{\theta}-1$,
\begin{eqnarray*}
& & \hspace{-0.2cm}
{\rm\bf Sm}^{[a]}_{\star,\sharp}(\varepsilon,\omega_{i}|_{i=1}^{N},\theta^{[0]})
= \lim_{\varepsilon \rightarrow 0}\frac{\left|\left| \theta_{v}^{[1]} - \theta^{[0]} \right|\right|_{\sharp}}
{\varepsilon}  \\
& & \hspace{-0.2cm}
{\rm\bf Sm}^{[r,k]}_{\star,,\sharp}(\varepsilon,\omega_{i}|_{i=1}^{N},\theta^{[0]})
= \lim_{\varepsilon \rightarrow 0} \frac{ \left|\left| \theta_{v}^{[k]} - \theta^{[0]} \right|\right|_{\sharp}}
{\left|\left| \theta_{v}^{[k+1]} - \theta^{[0]} \right|\right|_{\sharp}}
\end{eqnarray*}
are called respectively $\star$ and $\sharp$-norm based absolute and relative sloppiness metric of the descriptor system  $\Sigma$ at the parameter vector value $\theta^{[0]}$ with respect to the frequencies $\omega_{i}|_{i=1}^{N}$.
\end{definition}

Significance of this definition in system identification is obvious, noting that the absolute sloppiness metric is in fact the maximum ratio between parameter variations of the descriptor system $\rm\bf\Sigma$ and its frequency response deviation, while the relative sloppiness metric is the ratio between parameter variations in two succeeding extreme directions that lead to the same magnitude of its frequency response deviations. In this definition, the frequency response deviations are restricted to be very small, meaning that they are valid only when the associated frequency response is very close to that of the descriptor system $\rm\bf\Sigma(\theta^{[0]})$. This, however, does not mean that the corresponding parameter variations also have a small magnitude. These sloppiness metrics therefore appear more appropriate in investigating existence of sloppy parameter combinations, compared with the Fisher information matrix adopted in \cite{whrtt2021} and \cite{tmbdms2015}, in which parameter variations are restricted to have a small magnitude.

Clearly, this definition can also be easily modified to situations in which time-domain data is used in system identification.

The following results are well-known in matrix analyzes and computations \cite{gv1989,hj1991}, which are helpful in getting an identifiability condition and recursively searching desirable frequency samples.

\begin{lemma}
Divide a matrix $A$ as $A = \left[ A_{1}^{T} \;  A_{2}^{T}\right]^{T}$ in which the submatrix  $A_{1}$ is not of full column rank (FCR). Then the matrix $A$ is of FCR, if and only if the matrix $A_{2}A_{1,r}^{\perp}$ is.
\label{lemma:0}
\end{lemma}

It is worthwhile to point out that when the matrix $A_{1}$ is of FCR, the matrix $A = \left[ A_{1}^{T} \;  A_{2}^{T}\right]^{T}$ is obviously of FCR for each  matrix $A_{2}$ with a compatible dimension. In this case, however, $A_{1,r}^{\perp}=0$, meaning that the matrix $A_{2}A_{1,r}^{\perp} = 0 $ is not of FCR. These imply that the above lemma is not applicable to this situation.

\begin{lemma}
Divide a matrix $A$ as $A = {\rm\bf col}\left\{ A_{1}, \;\;  A_{2}\right\}$. Assume that the matrices $Z$ and $W$ are respectively an orthogonal basis for the right null space of the matrix $A_{1}$ and $A_{2}Z$. Then the matrix $ZW$ is an orthogonal basis for the right null space of the matrix $A$.
\label{lemma:0-orth}
\end{lemma}

\begin{lemma}
Let $A$, $B$ and $C$ be some prescribed complex matrices with compatible dimensions. Then there exists a matrix $X$ satisfying $AXB = C$, if and only if
\begin{displaymath}
(I-AA^{\dag})C = 0, \hspace{0.5cm} C(I-B^{\dag}B)=0
\end{displaymath}
When these two conditions are satisfied simultaneously, all the matrices satisfying $AXB = C$ can be expressed as follows,
\begin{displaymath}
X = A^{\dag} Y B^{\dag} + Z -A^{\dag}A Z BB^{\dag}
\end{displaymath}
in which $Z$ is an arbitrary matrix with an appropriate dimension.
\label{lemma:1}
\end{lemma}

Denote the SVDs of the matrices $A$ and $B$ respectively as follows,
\begin{eqnarray*}
& & A = \left[ U_{A,1}\;\; U_{A,2} \right]\left[\begin{array}{cc} \Sigma_{A} & 0 \\ 0 & 0 \end{array}\right] \left[ V_{A,1}\;\; V_{A,2} \right]^{H}  \\
& & B = \left[ U_{B,1}\;\; U_{B,2} \right]\left[\begin{array}{cc} \Sigma_{B} & 0 \\ 0 & 0 \end{array}\right] \left[ V_{B,1}\;\; V_{B,2} \right]^{H}
\end{eqnarray*}
in which the zero matrices have dimensions compatible with those of the matrices $\Sigma_{A}$ and $\Sigma_{B}$, that are in general different from each other. Then on the basis of the definition of the Moore-Penrose inverse, direct matrix operations show that the conclusions of the above lemma can be further expressed as that, there exists a solution to the equation $AXB = C$, if and only if $U^{H}_{A,2} C = 0$ and $C V_{B,2} = 0$.
In addition, when these two requirements are met, all the solutions to this equation can be parameterized as
\begin{displaymath}
X = V_{A,1} \Sigma^{-1}_{A} U^{H}_{A,1} Y V_{B,1} \Sigma^{-1}_{B} U^{H}_{B,1} + Z -V_{A,1} V^{H}_{A,1} Z U_{B,1} U^{H}_{B,1}
\end{displaymath}
with $Z$ being an arbitrary matrix of a compatible dimension.

When the matrices $A$, $B$ and $C$ are replaced by MVPs, similar conclusions can be obtained using their Smith-McMillan forms. The details are omitted due to their obviousness and space considerations.

\section{Parameter Identifiability}\label{section:si}

This section investigates identifiability verifications for the descriptor system $\rm\bf\Sigma$ of Section \ref{section:pfp} from its frequency response at finitely many frequencies. For this purpose, define TFMs $G_{zu}(\lambda)$, $G_{zv}(\lambda)$, $G_{yu}(\lambda)$ and $G_{yv}(\lambda)$ respectively as
\begin{eqnarray}\label{tfms}
	\left[\begin{array}{cc}
		G_{yu}(\lambda) & G_{yv}(\lambda)\\
		G_{zu}(\lambda) & G_{zv}(\lambda)
	\end{array}\right]
	& = &
	\left[\begin{array}{cc}
		D_{yu} & D_{yv}\\
		D_{zu} & D_{zv}
	\end{array}\right]
	+
	\left[\begin{array}{c}
		C_{yx}\\C_{zx}
	\end{array}\right]\times  \nonumber\\
& &	
	[\lambda E -A_{xx}]^{-1}
	\left[\begin{array}{cc}
		B_{xu} & B_{xv}
	\end{array}\right]
\end{eqnarray}
in which $\lambda$ stands for the Laplace/$\mathcal{Z}$ transformation variable $s$/$z$ for a continuous/discrete time descriptor system. Recall that the descriptor system $\rm\bf\Sigma$ is assumed to be regular for each $\theta \in {\rm\bf \Theta}$ which contains $0$ as its element. We have that the matrix pencil $\lambda E -A_{xx}$ is invertible. Hence, the above TFMs are well defined.

On the other hand, direct matrix manipulations show that,
\begin{eqnarray}
& & det\left(\lambda E - A(\theta)\right)\times det\left\{I_{m_v} - P(\theta) D_{zv}\right\} \nonumber\\
&=& det\left\{\lambda
\left[\begin{array}{cc}
		E & 0 \\ 0 & 0
\end{array}\right] -
\left[\begin{array}{cc}
		\! A_{xx} & B_{xv} P(\theta) \\
		C_{zx} & D_{zv} P(\theta) \!-\! I_{m_z}
	\!\end{array}\right] \right\} \nonumber\\
&=& det\left(\lambda E - A_{xx}\right)\times det\left\{I_{m_z} - G_{zv}(\lambda)P(\theta) \right\} \nonumber\\
&=& det\left(\lambda E - A_{xx}\right)\times det\left\{I_{m_v} - P(\theta) G_{zv}(\lambda)\right\}
\label{eqn:regularity}
\end{eqnarray}
It can therefore be declared that under the regularity and well-posedness assumptions A1) and A2), the inverses of both the TFMs $I_{m_v}- P(\theta) G_{zv}(\lambda)$ and $I_{m_z}- G_{zv}(\lambda) P(\theta)$ are also well defined. Based on these observations, it can be straightforwardly shown through some algebraic operations that, the TFM of the descriptor system $\rm\bf\Sigma$ from its external input $u(t)$ to its external output $y(t)$, that is, $H(\lambda,\theta)$, satisfies the following equalities,
\begin{eqnarray}
& & \hspace*{-1.5cm}H(\lambda,\theta)= D(\theta) + C(\theta)\left[\lambda E - A(\theta)\right]^{-1} B(\theta)  \nonumber \\
& &\hspace*{-0.4cm} = G_{yu}(\lambda)+G_{yv}(\lambda)[I_{m_v}- P(\theta) G_{zv}(\lambda)]^{-1} P(\theta) G_{zu}(\lambda)
\label{eqn:tfm}
\end{eqnarray}

In addition, the following conclusions can be established.

\begin{lemma}\label{lemma:3}
Assume that the descriptor system $\rm\bf\Sigma$ satisfies Assumptions A1 and A2)). Then its parameter vector is globally identifiable, if and only if there exist a finite integer $N$ and a set of distinct frequency samples $\omega_{i}|_{i=1}^{N}$, such that for every two different parameter vectors $\theta$ and $\widetilde{\theta}$ in the set ${\rm\bf\Theta}$, $H(j\omega_{i},\theta) =  H(j\omega_{i},\widetilde{\theta})$, $i=1,2,\cdots,N$, implies that $\theta = \widetilde{\theta}$.
\end{lemma}

While the above lemma gives a frequency domain criterion for the parameter identifiability of the descriptor system $\rm\bf\Sigma$, it is still not very clear how to verify the associated conditions. In addition, it may be desirable to know the minimum number $N$ for the frequency samples, as well as the values of these frequency samples, such that the associated system frequency response samples are able to distinguish system parameters of the descriptor system $\rm\bf\Sigma$.

To develop a computationally feasible condition, the following results are at first introduced.

Let $\Pi$ be a complex matrix. Assume that its singular value decomposition (SVD) takes the following form
\begin{displaymath}
\Pi = \left[ U_{\Pi,1}\;\; U_{\Pi,2} \right]\left[\begin{array}{cc} \Sigma_{\Pi} & 0 \\ 0 & 0 \end{array}\right] \left[ V_{\Pi,1}\;\; V_{\Pi,2} \right]^{H}
\end{displaymath}
in which both the matrix $\left[ U_{\Pi,1}\;\; U_{\Pi,2} \right]$ and the matrix $\left[ V_{\Pi,1}\;\; V_{\Pi,2} \right]$ are unitary, while the matrix $\Sigma_{\Pi}$ is diagonal with each of its diagonal elements greater than zero. In addition, the zero matrices usually have different dimensions.
When this matrix is of FCR over the complex ring, it can be straightforwardly shown from matrix analyzes \cite{hj1991} that the matrix $V_{\Pi,2}$ is empty. Under such a situation, the matrix $V_{\Pi,1}$ is simply written as $V_{\Pi}$.

Denote the real and imaginary parts of the matrix $\Pi$ respectively by $\Pi_{r}$ and $\Pi_{j}$. Then it can be directly shown through matrix manipulations that the complex matrix $\Pi$ is of FCR over the complex field, if and only if the following real matrix
\begin{displaymath}
\left[\begin{array}{rr} \Pi_{r} & -\Pi_{j} \\ \Pi_{j} & \Pi_{r} \end{array} \right]
\end{displaymath}
is of FCR over the real field, noting that for an arbitrary complex vector $\alpha$ with a compatible dimension, the following equality is always valid
\begin{displaymath}
\Pi \alpha = (\Pi_{r} \alpha_{r}  -\Pi_{j} \alpha_{j}) + j (\Pi_{j} \alpha_{r} +  \Pi_{r} \alpha_{j})
\end{displaymath}
in which $\alpha_{r}$ and $\alpha_{j}$ denote respectively the real and imaginary parts of the vector $\alpha$.
Hence when the complex matrix $\Pi$ is of FCR over the complex field, the following matrix is well defined
\begin{displaymath}
\left[\!\!\begin{array}{cc} \Pi_{r} & -\Pi_{j} \end{array} \!\!\right]
\left(\left[\!\!\begin{array}{rr} \Pi_{r} & -\Pi_{j} \\ \Pi_{j} & \Pi_{r} \end{array} \!\!\right]^{T}\left[\!\!\begin{array}{rr} \Pi_{r} & -\Pi_{j} \\ \Pi_{j} & \Pi_{r} \end{array} \!\!\right]\right)^{-1}\!\!
\left[\!\!\begin{array}{cc} \Pi_{r} & -\Pi_{j} \end{array} \!\!\right]^{T}
\end{displaymath}

On the basis of these symbols and results, the following conclusions can be obtained, which are quite helpful in the derivations of a computationally attractive condition for frequency response based identifiability.

\begin{lemma}\label{lemma:4}
Assume that a complex matrix $\Pi$ is of FCR. Then
\begin{eqnarray}
& &\hspace*{-0.2cm} I \!-\! \left[\!\!\begin{array}{cc} \Pi_{r} & -\Pi_{j} \end{array} \!\!\right]
\left(\left[\!\!\begin{array}{rr} \Pi_{r} & -\Pi_{j} \\ \Pi_{j} & \Pi_{r} \end{array} \!\!\right]^{T} \!\! \left[\!\!\begin{array}{rr} \Pi_{r} & -\Pi_{j} \\ \Pi_{j} & \Pi_{r} \end{array} \!\!\right]\right)^{-1}\!\!
\!\!\left[\!\!\begin{array}{cc} \Pi_{r} & -\Pi_{j} \end{array} \!\! \right]^{T} \nonumber \\
& &\hspace*{-0.4cm} = \left[\!\!\begin{array}{cc} U_{\Pi,2r} & U_{\Pi,2j} \end{array} \!\!\right]
\left[\!\!\begin{array}{cc} U_{\Pi,2r} & U_{\Pi,2j} \end{array} \!\!\right]^{T}
\end{eqnarray}
in which $U_{\Pi,2r}$ and $U_{\Pi,2j}$ stand respectively for the real and imaginary parts of the matrix $U_{\Pi,2}$.
\end{lemma}

For each TFM $G_{\#\star}(\lambda)$ with $\#,\star$ being $u$, $y$, $z$ or $v$, denote its Smith-McMillan form by
\begin{displaymath}
\left[ U_{\#\star, 1}(\lambda)\;\;U_{\#\star, 2}(\lambda) \right]
\left[\begin{array}{cc} \Sigma_{\#\star}(\lambda) & 0 \\ 0 & 0 \
\end{array}\right] \left[ V_{\#\star, 1}(\lambda)\;\;V_{\#\star, 2}(\lambda) \right]^{T}
\end{displaymath}
in which both $\left[ U_{\#\star, 1}(\lambda)\;\;U_{\#\star, 2}(\lambda) \right]$ and $\left[ V_{\#\star, 1}(\lambda)\;\;V_{\#\star, 2}(\lambda) \right]$ are unimodular MVPs, while $\Sigma_{\#\star}(\lambda)$ is a diagonal matrix with each of its diagonal elements being a finite degree rational function that is not identically equal to zero. The zero matrices of the aforementioned Smith-McMillan form usually have different dimensions, which are omitted for brevity. Moreover, partition the inverse of these unimodular MVPs compatible with the dimensions of the MVFs $\Sigma_{\#\star}(\lambda)$ as follows,
\begin{displaymath}
\left[ \clubsuit_{\#\star, 1}(\lambda)\;\;\clubsuit_{\#\star, 2}(\lambda) \right]^{-1}
= \left[ \begin{array}{c} \clubsuit^{[inv,1]}_{\#\star}(\lambda)  \\
\clubsuit^{[inv,2]}_{\#\star}(\lambda) \end{array} \right]
\end{displaymath}
in which $\clubsuit = U$ or $V$, while $\#,\star = u$, $y$, $z$ or $v$.

With these symbols, define a MVF $\Pi(\theta,\lambda)$ using the TFMs  $G_{yv}(\lambda)$ and $G_{zv}(\lambda)$ as follow,
\begin{equation}
\Pi(\lambda,\theta) = \left[ I - P(\theta) G_{zv}(\lambda) \right] V_{yv}^{[inv,2]T}(\lambda)
\label{eqn:pi}
\end{equation}
In addition, denote the real and imaginary parts of the associated complex matrix $\Pi(j\omega,\theta)$ respectively by $\Pi_{r}(\omega,\theta)$ and $\Pi_{j}(\omega,\theta)$. Moreover, define real matrices $\overline{\Pi}_{r}(\omega,\theta)$, $\overline{\Pi}_{j}(\omega,\theta)$ and $\Xi(\omega,\theta)$ respectively as
\begin{eqnarray*}
& & \overline{\Pi}_{r}(\omega,\theta) = \left[ \Pi_{r}(\omega,\theta)\;\; -\Pi_{j}(\omega,\theta) \right] \\
& & \overline{\Pi}_{j}(\omega,\theta) = \left[ \Pi_{j}(\omega,\theta)\;\; \Pi_{r}(\omega,\theta) \right] \\
& & \Xi(\omega,\theta) = \left( \overline{\Pi}_{r}(\omega,\theta) \overline{\Pi}_{j,r}^{\perp}(\omega,\theta) \right)_{l}^{\perp}
\end{eqnarray*}
Furthermore, express the SVD of the complex matrix $\Pi(j\omega,\theta)$ as
\begin{eqnarray}
\Pi(j\omega,\theta) \!\!\!\!&=&\!\!\!\! \left[ U_{\Pi,1}(j\omega,\theta)\;\; U_{\Pi,2}(j\omega,\theta) \right]\left[\begin{array}{cc} \Sigma_{\Pi}(\omega,\theta) & 0 \\ 0 & 0 \end{array}\right] \times \nonumber \\
& &  \!\!\!\!\left[ V_{\Pi,1}(j\omega,\theta)\;\; V_{\Pi,2}(\omega,\theta) \right]^{H}
\label{eqn:svd-2}
\end{eqnarray}

On the other hand, define a matrix $\Psi$ using the matrices $P_{i}|_{i=1}^{q}$ of Eq.(\ref{eqn:ParVec}) as follows,
\begin{equation}
\Psi = \left[ {\rm\bf vec}(P_{1})\;\; {\rm\bf vec}(P_{2})\;\; \cdots \;\; {\rm\bf vec}(P_{q}) \right]
\label{eqn:par-vec-1}
\end{equation}
Moreover, assume its SVD takes the following form
\begin{equation}
\Psi = \left[ U_{\Psi,1}\;\; U_{\Psi,2} \right]\left[\begin{array}{cc} \Sigma_{\Psi} & 0 \\ 0 & 0 \end{array}\right] \left[ V_{\Psi,1}\;\; V_{\Psi,2} \right]^{T}
\label{eqn:par-vec-2}
\end{equation}
in which both the matrix $\left[ U_{\Psi,1}\;\; U_{\Psi,2} \right]$ and the matrix $\left[ V_{\Psi,1}\;\; V_{\Psi,2} \right]$ are orthogonal, while the matrix $\Sigma_{\Psi}$ is diagonal with each of its diagonal elements being greater than zero. Once again, the associated zero matrices in the above expression usually have different dimensions.

Based on these decompositions, the following necessary and sufficient condition is obtained for the identifiability of the descriptor system $\rm\bf\Sigma$ at a particular parameter vector value using its frequency response at a set of finitely many frequencies.

\begin{theorem}\label{theo:1}
The descriptor system $\rm\bf \Sigma$ is identifiable, only when the matrix $\Psi$ is of FCR. When this condition is satisfied, assume that the TFM $G_{zu}(\lambda)$ is of FNRR. Then the descriptor system $\rm\bf\Sigma$ is globally identifiable at a specific parameter vector  $\theta^{[0]} \in {\rm\bf\Theta}$, if there exists a frequency sample  $\omega_{1}$, at which the matrix $\overline{\Pi}_{j}(\omega_{1},\theta^{[0]})$ is of FCR. Otherwise, the descriptor system $\rm\bf\Sigma$ is globally identifiable at $\theta^{[0]} \in {\rm\bf\Theta}$, if and only if there exists a finite integer $N$ and a set of distinct frequencies $\omega_{i}|_{i=1}^{N}$, such that both the matrix
$\overline{\Pi}_{r}(\omega_{1},\theta^{[0]}) \overline{\Pi}_{j,r}^{\perp}(\omega_{1},\theta^{[0]})$
and the matrix $\Upsilon(\omega_{i}|_{i=1}^{N},\theta^{[0]})$ are of FCR, in which the matrix $\Upsilon(\omega_{i}|_{i=1}^{N},\theta^{[0]})$ is defined as follows,
\begin{equation}\label{eqn:theo1}
\Upsilon(\omega_{i}|_{i=1}^{N},\theta^{[0]}) =
\left[\begin{array}{c}
U^{T}_{\Psi,2} \\
I \otimes \left[ \begin{array}{c} \Xi(\omega_{1}, \theta^{[0]}) \\
\left[ U_{\Pi,2r}(\omega_{2},\theta^{[0]}) \;\; U_{\Pi,2j}(\omega_{2},\theta^{[0]}) \right]^{T} \\
\left[ U_{\Pi,2r}(\omega_{3},\theta^{[0]}) \;\; U_{\Pi,2j}(\omega_{3},\theta^{[0]}) \right]^{T} \\
\vdots \\
\left[ U_{\Pi,2r}(\omega_{N},\theta^{[0]}) \;\; U_{\Pi,2j}(\omega_{N},\theta^{[0]}) \right]^{T} \end{array}\right] \end{array}\right]
\end{equation}
Here, for each $i=2,3,\cdots,N$, $U_{\Pi,2r}(\omega_{i},\theta^{[0]})$ and $U_{\Pi,2j}(\omega_{i},\theta^{[0]})$ stand respectively for the real and imaginary parts of the complex matrix $U_{\Pi,2}(j\omega_{i},\theta^{[0]})$ defined through the SVD of the complex matrix $\Pi(j\omega_{i},\theta^{[0]})$.
\end{theorem}

From Lemma \ref{lemma:0}, it is clear that when the frequencies $\omega_{i}|_{i=1}^{N}$ are given, the conditions of Eq.(\ref{eqn:theo1}), that is,  whether or not the matrix $\Upsilon(\omega_{i}|_{i=1}^{N},\theta^{[0]})$ is of FCR, can be verified recursively. This is quite appealing, especially when the frequency number $N$ is large and/or the system has a high dimension. The latter usually leads to a high dimensional matrix $U_{\Pi,2}(j\omega,\theta^{[0]})$.

\section{Determination of Desirable Frequency Samples}\label{section:ddfs}

Lemma \ref{lemma:3} and Theorem \ref{theo:1} of the previous section make it possible to verify parameter identifiability of the descriptor system $\rm\bf\Sigma$ at a particular parameter vector using its frequency response at finitely many frequencies. The required number of the frequencies, however, is in general large, especially when the TFM of the descriptor system $\rm\bf\Sigma$ has a high degree, which is the general situation when a large scale NDS is under investigation.

In this section, we study possibilities of using less frequency response information in parameter identifiability verification for the descriptor system $\rm\bf\Sigma$, as well as determination of the desirable frequencies. To reduce the number of the required frequencies, the following results are at first introduced.

\begin{lemma}\label{lemma:5}
For an arbitrary proper TFM $\Pi(\lambda)$, let $N(\lambda)D^{-1}(\lambda)$ denote one of its right coprime factorizations over the MVP ring. Denote the real and imaginary parts of $\Pi(j\omega)$ respectively by $\Pi_{r}(\omega)$ and $\Pi_{j}(\omega)$, while those of $N(j\omega)$ by $N_{r}(\omega)$ and $N_{j}(\omega)$. Then
\begin{eqnarray}
& & \left(\left[ \Pi_{r}(\omega)\;\; -\Pi_{j}(\omega) \right] \left[ \Pi_{j}(\omega)\;\; \Pi_{r}(\omega) \right]_{r}^{\perp}\right)_{l}^{\perp} \nonumber\\
&=&
\left(\left[ N_{r}(\omega)\;\; -N_{j}(\omega) \right] \left[ N_{j}(\omega)\;\; N_{r}(\omega) \right]_{r}^{\perp}\right)_{l}^{\perp}
\end{eqnarray}
\end{lemma}

Assume that $N(\lambda)$ is a real MVP and $N(\lambda)=\sum_{i=0}^{m}N_{i}\lambda^{i}$. Define an MVP $\overline{N}(\lambda)$ as
\begin{displaymath}
\overline{N}(\lambda) = \sum_{i=0}^{m}(-1)^{\lfloor i/2 \rfloor} N_{i}\lambda^{i}
\end{displaymath}
in which $\lfloor i/2 \rfloor$ denotes the greatest integer that is not greater than $i/2$. Denote by $\overline{N}_{j}(\lambda)$ and $\overline{N}_{r}(\lambda)$ the MVPs consisting of respectively only the odd and even terms of the MVP $\overline{N}(\lambda)$.

Assume that the MVP $\left[\overline{N}_{j}(\lambda)\;\; \overline{N}_{r}(\lambda)\right]$ takes the following Smith form
\begin{eqnarray*}
& & \hspace*{-1.0cm} \left[\overline{N}_{j}(\lambda)\;\; \overline{N}_{r}(\lambda)\right]
=
\left[ U_{\overline{N},1}(\lambda)\;\; U_{\overline{N},2}(\lambda)\right]
\left[ \begin{array}{cc} \Sigma_{\overline{N}}(\lambda) & 0  \\
0 & 0 \end{array}\right]\times\\
& & \hspace*{4.5cm}
\left[ V_{\overline{N},1}(\lambda)\;\; V_{\overline{N},2}(\lambda)\right]^{T}
\end{eqnarray*}
Then there exist MVPs $V^{[inv,1]}_{\overline{N}}(\lambda)$ and $V^{[inv,2]}_{\overline{N}}(\lambda)$ with their dimensions compatible with those of the MVPs $V_{\overline{N},1}(\lambda)$ and $V_{\overline{N},2}(\lambda)$, such that
\begin{displaymath}
\left[ V_{\overline{N},1}(\lambda)\;\; V_{\overline{N},2}(\lambda)\right]^{-1} =
\left[ \begin{array}{c} V^{[inv,1]}_{\overline{N}}(\lambda)  \\
V^{[inv,2]}_{\overline{N}}(\lambda) \end{array}\right]
\end{displaymath}

Moreover, assume that the MVP $\left[\overline{N}_{r}(\lambda)\;\; -\overline{N}_{j}(\lambda)\right]\times$ $V^{[inv,2]T}_{\overline{N}}(\lambda)$ takes the following Smith form
\begin{eqnarray*}
& & \hspace*{-0.1cm} \left[\overline{N}_{r}(\lambda)\;\; -\overline{N}_{j}(\lambda)\right] V^{[inv,2]T}_{\overline{N}}(\lambda) \\
& & \hspace*{-0.5cm}=
\left[ U_{\widetilde{N},1}(\lambda)\;\; U_{\widetilde{N},2}(\lambda)\right]
\left[ \begin{array}{cc} \Sigma_{\widetilde{N}}(\lambda) & 0  \\
0 & 0 \end{array}\right]
\left[ V_{\widetilde{N},1}(\lambda)\;\; V_{\widetilde{N},2}(\lambda)\right]^{T}
\end{eqnarray*}
and
\begin{displaymath}
\left[ U_{\widetilde{N},1}(\lambda)\;\; U_{\widetilde{N},2}(\lambda)\right]^{-1} =
\left[ \begin{array}{c} U^{[inv,1]}_{\widetilde{N}}(\lambda)  \\
U^{[inv,2]}_{\widetilde{N}}(\lambda) \end{array}\right]
\end{displaymath}
in which $U^{[inv,1]}_{\widetilde{N}}(\lambda)$ and $U^{[inv,2]}_{\widetilde{N}}(\lambda)$ are MVPs with their dimensions being consistent with those of the MVPs $U_{\widetilde{N},1}(\lambda)$ and $U_{\widetilde{N},2}(\lambda)$.

With these symbols, the following conclusions can be established from Lemma  \ref{lemma:5}.

\begin{theorem}\label{theo:3}
Assume that $\Pi(\lambda)$ is a proper and stable TFM with a finite degree. Let $N(\lambda)D^{-1}(\lambda)$ denote one of its right coprime factorizations over the MVP ring. Denote the real and imaginary parts of its frequency response  respectively by $\Pi_{r}(\omega)$ and $\Pi_{j}(\omega)$. Then for each $\omega \in {\cal R}$, the matrix $U^{[inv,2]}_{\widetilde{N}}(\omega)$ is uniquely determined by the following expression, modulo an orthogonal transformation,
\begin{equation}
U^{[inv,2]}_{\widetilde{N}}(\omega) =
\left(\left[ \Pi_{r}(\omega)\;\; -\Pi_{j}(\omega) \right] \left[ \Pi_{j}(\omega)\;\; \Pi_{r}(\omega) \right]_{r}^{\perp}\right)_{l}^{\perp}
\end{equation}
\end{theorem}

When a MVF $\Pi(\lambda)$ is not proper, it can always be decomposed into the following form
\begin{displaymath}
\Pi(\lambda) = \Pi_{ac}(\lambda) + \Pi_{c}(\lambda)
\end{displaymath}
in which $\Pi_{ac}(\lambda)$ is an MVP, while $\Pi_{c}(\lambda)$ is a proper TFM. Then over the MVP ring, there exists a right coprime factorization for the proper TFM $\Pi_{c}(\lambda)$, denote it by $N_{c}(\lambda)D_{c}^{-1}(\lambda)$. With this decomposition, the MVF $\Pi(\lambda)$ can be equivalently written as
\begin{displaymath}
\Pi(\lambda) = \left[\Pi_{ac}(\lambda)D_{c}(\lambda) + N_{c}(\lambda)\right] D_{c}^{-1}(\lambda)
\end{displaymath}

With this expression for the MVF $\Pi(\lambda)$, it is obvious from the proofs Lemma \ref{lemma:5} and Theorem \ref{theo:3} that, their conclusions can still be applied to the situation in which this MVF is not proper. The only required modifications are to replace the associated MVPs $N(\lambda)$ and $D(\lambda)$ respectively with the MVPs $\Pi_{ac}(\lambda)D_{c}(\lambda) + N_{c}(\lambda)$ and $D_{c}(\lambda)$.

Note that for a MVP, its Smith form has an analytic expression, as well as the inverse of an unimodular MVP. Results of Theorem \ref{theo:3} makes it clear that when the value of the parameter vector $\theta$ is given, an explicit expression is available for the matrix $\Xi(\omega,\theta)$ defined in the previous section, and therefore enables an optimal search of the frequency $\omega_{1}$ required in the identifiability condition of Theorem \ref{theo:1}.

On the other hand, from the definition of the MVP $V_{yv,1}(\lambda)$, it is obvious that it is of FCR for each $\lambda \in {\cal C}$. Note that the TFM $I - P(\theta)G_{zv}(\lambda)$ is guaranteed to be invertible for every $\theta \in {\rm\bf\Theta}$ by the adopted regularity and well-posedness assumptions A1) and A2). It can therefore be declared that the MVF $\left[I - P(\theta)G_{zv}(j\omega)\right]^{-1}V_{yv,1}(j\omega)$ is of FCR for each $\omega \in {\cal R}$ whenever $\theta$ belongs to the set ${\rm\bf\Theta}$. These observations imply that for each $\theta \in {\rm\bf\Theta}$, there exist an inner MVF $U_{in}(\lambda,\theta)$ and an outer MVF $U_{out}(\lambda,\theta)$, such that
\begin{equation}
\left[ I - P(\theta)G_{zv}(-\lambda) \right]^{-T}V_{yv,1}(-\lambda) = U_{in}(\lambda,\theta)U_{out}(\lambda,\theta)
\label{eqn:in-out}
\end{equation}
for the continuous time descriptor system $\rm\bf\Sigma$ \cite{zdg1996}. When the descriptor system $\rm\bf\Sigma$ is of discrete time, this equation must be modified into the following form,
\begin{displaymath}
\left[ I - P(\theta)G_{zv}(1/\lambda) \right]^{-T}V_{yv,1}(1/\lambda) = U_{in}(\lambda,\theta)U_{out}(\lambda,\theta)
\end{displaymath}

On the basis of this factorization, we have the following conclusions about the SVD of the frequency response of the MVF $\Pi(\lambda,\theta)$.

\begin{theorem}\label{theo:4}
Assume that the MVF $\Pi(\lambda,\theta)$ of Eq.(\ref{eqn:pi}) is proper. Then for each $\theta \in {\rm\bf\Theta}$ and each $\omega \in {\cal R}$, up to an unitary transformation, the MVF $U_{\Pi,2}(j\omega,\theta)$ defined in the SVD of the MVF $\Pi(j\omega,\theta)$ through  Eq.(\ref{eqn:svd-2}) is uniquely determined by
\begin{equation}
U_{\Pi,2}(j\omega,\theta) = U_{in}(j\omega,\theta)
\end{equation}
\end{theorem}

When the MVF $\Pi(\lambda,\theta)$ of Eq.(\ref{eqn:pi}) is not proper, a slight modification can enable the application of the above theorem. In particular, through simply multiplying the  MVF $\Pi(\lambda,\theta)$ from left by a factor $(\lambda + a )^{-m}$ with some appropriate $a$ and $m$, this MVF can be made proper. Note that this factor changes only the singular values of the associate matrix, and does not have any effects on the required unitary matrices. It is safe to declare that results of Theorem \ref{theo:4} is able to give an explicit expression for the matrix $U_{\Pi,2}(j\omega,\theta)$ of Eq.(\ref{eqn:svd-2}).

Note that for an arbitrary TFM whose frequency response is of FCR at each frequency, its inner-outer factorization has an analytic expression using its state space realization. Theorem \ref{theo:4} makes it clear that for each frequency, both the matrix $U_{\Pi,2r}(\omega,\theta)$ and the matrix $U_{\Pi,2j}(\omega,\theta)$ have an analytic expression also, provided that the value of the parameter vector $\theta$ is known.

Recall that $U^{T}_{\Psi,2} U_{\Psi,1} = 0$ and $U^{T}_{\Psi,1} U_{\Psi,1} = I$.  These mean that the column vectors of the matrix $U_{\Psi,1}$ consists of an orthogonal basis for the right null space of the matrix $U^{T}_{\Psi,2}$. On the basis of Lemma \ref{lemma:0-orth}, as well as Theorems \ref{theo:1}, \ref{theo:3} and \ref{theo:4}, the following algorithm can be constructed for recursively searching the desirable frequencies, at which the frequency response of descriptor system $\rm\bf\Sigma$ is able to uniquely determine the value of its parameter vector $\theta$, provided that it is identifiable.

\renewcommand{\labelenumi}{\rm\bf S\arabic{enumi})}
\renewcommand{\labelenumii}{\rm\bf s\arabic{enumi})}

\begin{enumerate}
\item Let $k=1$. Find a frequency sample $\omega_{1}$, such that the matrix $\overline{\Pi}_{r}(\omega_{1},\theta^{[0]}) \overline{\Pi}_{j,r}^{\perp}(\omega_{1},\theta^{[0]})$ is of FCR, while the matrix $\left\{I \otimes \Xi(\omega_{1}, \theta^{[0]}) \right\} U_{\Psi,1}$ has the maximum number of independent columns.
\item Find an orthogonal basis for the right null space of the matrix $\left\{I \otimes \Xi(\omega_{1}, \theta^{[0]}) \right\} U_{\Psi,1}$. Denote it by $Z(\omega_{1})$.

    If the  matrix $Z(\omega_{1})$ is equal to a zero vector, the descriptor system $\rm\bf\Sigma$ is globally identifiable at $\theta^{[0]}$, if and only if its frequency responses are  available at $\omega_{1}$ . Finish the computations.

    Otherwise, go to the next step.

\item Find a frequency sample $\omega_{k+1}$, such that the matrix $\left\{I \otimes \left[ U_{\Pi,2r}(\omega_{k+1},\theta^{[0]}) \;\; U_{\Pi,2j}(\omega_{k+1},\theta^{[0]}) \right]^{T} \right\} Z(\omega_{k})$ has the maximum number of independent columns.

\item Find an orthogonal basis for the right null space of the matrix $\left\{I \otimes \left[ U_{\Pi,2r}(\omega_{k+1},\theta^{[0]}) \;\; U_{\Pi,2j}(\omega_{k+1},\theta^{[0]}) \right]^{T} \right\} Z(\omega_{k})$. Denote it by $\widetilde{Z}(\omega_{k+1})$.

    If the  matrix $\widetilde{Z}(\omega_{k+1})$ is equal to a zero vector, the descriptor system $\rm\bf\Sigma$ is globally identifiable at $\theta^{[0]}$, if and only if its frequency responses are available at $\omega_{i}|_{i=1}^{k+1}$ . Finish the computations.

    If the matrix $\widetilde{Z}(\omega_{k+1})$ is invertible, the descriptor system $\rm\bf\Sigma$ is not identifiable at $\theta^{[0]}$. Finish the computations.

    Otherwise, go to the next step.

\item Assign $k$ as $k+1$. Let $Z(\omega_{k+1}) = Z(\omega_{k}) \widetilde{Z}(\omega_{k+1})$. Go to Step {\rm\bf S3)}.

\end{enumerate}

Note that Theorems \ref{theo:3} and \ref{theo:4} give respectively an analytic expression for the MVFs $\Xi(\omega, \theta)$ and $U_{\Pi,2}(j\omega,\theta)$ for each $\omega \in {\cal R}$ and each $\theta \in {\rm\bf\Theta}$. The optimizations of Steps S1) and S3) in the above algorithm can be efficiently performed through getting the Smith form of the associated MVP or the Smith-McMillan form of the associated MVF.

On the other hand, note that the right null space of the matrix $U^{T}_{\Psi,2}$ has a finite dimension. It is certain that the above algorithm finishes in finite steps, noting that if in a computation recursion, the number of columns of the matrix $Z(\omega_{k+1})$ has not been reduced, which is equivalent to that the matrix $\widetilde{Z}(\omega_{k+1})$ is invertible, then the descriptor system $\rm\bf\Sigma$ is not globally identifiable at this particular parameter vector value $\theta^{[0]}$, and the computations will be finished.

In addition, while the results of Theorems \ref{theo:3} and \ref{theo:4} are unique only respectively up to an orthogonal transformation and an unitary transformation, it is obvious from Theorem \ref{theo:1} that these transformations do not affect the validness of the aforementioned frequency search algorithm.

Note that a TFM is of FNCR, if and only if its transpose is of FNRR. The results of the previous section and this section are also applicable to situations in which the TFM $G_{yv}(\lambda)$ is of FNCR. The only required modifications ar to take a transpose of the TFM $H(\lambda,\theta)$ of the descriptor system $\rm\bf\Sigma$.

\section{Existence of Sloppy Parameter Combinations}

In frequency domain system identification, frequency responses of a system are usually estimated from time domain input/output data. This means that there  usually exist estimation errors in the adopted data \cite{Ljung1999,ps2001}. On the other hand, when there is a large number of parameters to be identified, there may exist the so-called sloppy parameter combinations, meaning that parameter variations along these directions can hardly be detected from data \cite{tmbdms2015,wcgbmbes2006}. Therefore, in addition to identifiability, it is also important to investigate whether or not there are some values for an identifiable parameter vector, along which system frequency responses are not very sensitive to its variations.

In this section, we study existence possibilities of a large value deviation of the parameter vector $\theta$ from a globally identifiable parameter vector value, say, $\theta^{[0]}$, that leads to only very small variations of the frequency responses of the descriptor system $\rm\bf\Sigma$. An explicit description is derived for all the possible parameter vector values, which constitute approximately a convex set. Analytic expressions are obtained from this approximation respectively for the absolute and relative sloppiness metrics.

It is now well known that under some weak conditions, estimation errors of the frequency responses of a system are asymptotically normal with their expectations being zero, and their real and imaginary parts are asymptotically independent of each other, while estimation errors at different frequencies are also asymptotically independent of each other \cite{Ljung1999,ps2001}. Note that the sum of the squares of several independent Gaussian random variables has a $\chi^{2}$ distribution. It therefore appears reasonable to assume that the sum of the magnitude square of (weighted) estimation errors of system frequency responses belongs to a bounded interval away from zero. That is, this weighted sum is in general greater than a positive number with a very high probability, if not with probability $1$, especially when the number of the sampled frequencies, that is, $N$, is large, and/or the TFM of the descriptor system $\rm\bf\Sigma$ has a high dimension which is the normal situation for a large scale NDS.

These arguments mean that it is essential to investigate existence of a large parameter value variations for the descriptor system $\rm\bf\Sigma$ under the situation in which the aforementioned sum is less than a small positive number. In other words, to have an estimate for its parameters with an acceptable precision, it is necessary that a small $\sqrt{\sum_{i=1}^{N}|| H(j\omega_{i},\widetilde{\theta})  - H(j\omega_{i},\theta)||_{F}^{2}}$ is resulted {\it only} from a small $||\widetilde{\theta} - \theta||_{2}$, in which $\theta,\widetilde{\theta} \in {\rm\bf\Theta}$, and $\omega_{i}|_{i=1}^{N}$ are some distinct frequencies.

To study this problem, let $\theta^{[0]}$ be a particular parameter vector belonging to the set ${\rm\bf\Theta}$. Define real MVFs $\Phi_{r}(\lambda,\theta^{[0]})$ and $\Phi_{l}(\lambda,\theta^{[0]})$ respectively as
\begin{eqnarray*}
& & \hspace*{-1.0cm}
\Phi_{r}(\lambda,\theta^{[0]}) =
\Sigma_{zu}^{-1}(\lambda) U_{zu}^{[inv,1]}(\lambda) \left[I -  G_{zv}(\lambda) P(\theta^{[0]}) \right] \\
& & \hspace*{-1.0cm} \Phi_{l}(\lambda,\theta^{[0]}) =
\left[I \!-\! P(\theta^{[0]}) G_{zv}(\lambda) \right]
V_{yv}^{[inv,1]T}(\lambda)\Sigma_{yv}^{-1}(\lambda)
\end{eqnarray*}
Moreover, for every $\omega \in {\cal R}$, denote the real and imaginary parts of the complex matrices $\Phi_{l}(j\omega,\theta^{[0]})$ and $\Phi_{r}(j\omega,\theta^{[0]})$ respectively by
$\Phi_{l,r}(\omega,\theta^{[0]})$, $\Phi_{l,j}(\omega,\theta^{[0]})$,
$\Phi_{r,r}(\omega,\theta^{[0]})$ and $\Phi_{r,j}(\omega,\theta^{[0]})$.  Furthermore, define real matrices $Q_{r}(\omega,\theta^{[0]})$ and $Q_{j}(\omega,\theta^{[0]})$ respectively as \small
\begin{eqnarray*}
& & \hspace*{-1.00cm}
Q_{r}(\omega,\theta^{[0]}) \!=\! \left[
\Phi_{r,r}^{T}(\omega,\theta^{[0]}) \otimes \Phi_{l,r}(\omega,\theta^{[0]}) -
\Phi_{r,j}^{T}(\omega,\theta^{[0]}) \otimes \Phi_{l,j}(\omega,\theta^{[0]}) \right. \\
& & \hspace*{0.50cm} \left.
-\Phi_{r,j}^{T}(\omega,\theta^{[0]}) \!\otimes\! \Phi_{l,r}(\omega,\theta^{[0]}) \!-\!
\Phi_{r,r}^{T}(\omega,\theta^{[0]}) \!\otimes\! \Phi_{l,j}(\omega,\theta^{[0]})
\right]T_{\!\! H}
 \\
& & \hspace*{-1.00cm}
Q_{l}(\omega,\theta^{[0]}) \!=\! \left[
\Phi_{r,j}^{T}(\omega,\theta^{[0]}) \otimes \Phi_{l,r}(\omega,\theta^{[0]}) +
\Phi_{r,r}^{T}(\omega,\theta^{[0]}) \otimes \Phi_{l,j}(\omega,\theta^{[0]}) \right. \\
& & \hspace*{0.70cm} \left.
\Phi_{r,r}^{T}(\omega,\theta^{[0]}) \!\otimes\! \Phi_{l,r}(\omega,\theta^{[0]}) \!-\!
\Phi_{r,j}^{T}(\omega,\theta^{[0]}) \!\otimes\! \Phi_{l,j}(\omega,\theta^{[0]})
 \right]T_{\!\! H}
\end{eqnarray*}
\normalsize
in which
\small
\begin{displaymath}
T_{\!\! H}=\left[\begin{array}{ccccccc}
I & 0 & 0 & 0 & \cdots & 0 & 0 \\
0 & 0 & I & 0 & \cdots & 0 & 0 \\
\vdots & \vdots & \vdots & \vdots & \ddots & \vdots & \vdots \\
0 & 0 & 0 & 0 & \cdots & I & 0 \\
0 & I & 0 & 0 & \cdots & 0 & 0 \\
0 & 0 & 0 & I & \cdots & 0 & 0 \\
\vdots & \vdots & \vdots & \vdots & \ddots & \vdots & \vdots \\
0 & 0 & 0 & 0 & \cdots & 0 & I \end{array}\right]
\end{displaymath}
\normalsize

Let $\omega_{i}|_{i=1}^{N}$ be a set of distinct frequencies. Define a matrix $\Omega(\omega_{i}|_{i=1}^{N},\theta^{[0]})$ similarly as the matrix $\Gamma(\omega_{i}|_{i=1}^{N},\theta^{[0]})$ of  Eq.(\ref{eqn:a28}) through the matrices $\Psi$, $Q_{r}(\omega_{i},\theta^{[0]})|_{i=1}^{N}$ and $Q_{j}(\omega_{i},\theta^{[0]})|_{i=1}^{N}$. Moreover, define matrices $S_{\!\! H}$ and $S_{\!\! A}$ respectively as follows,
\begin{eqnarray*}
& & \hspace*{-0.8cm}
S_{\!\! H}
=
\left\{\Gamma_{l}^{\perp}(\omega_{i}|_{i=1}^{N},\theta^{[0]})
\Omega(\omega_{i}|_{i=1}^{N},\theta^{[0]})\right\}_{r}^{\perp} \\
& & \hspace*{-0.8cm}
S_{\!\! A}
= \Gamma^{\dag}(\omega_{i}|_{i=1}^{N},\theta^{[0]}) \Omega(\omega_{i}|_{i=1}^{N},\theta^{[0]}) S_{\!\! H}(\omega_{i}|_{i=1}^{N},\theta^{[0]})
\end{eqnarray*}
Denote the number of the matrix $S_{\!\! H}$ by $n_{s}$. Partition the matrices $S_{\!\! H}$ and $S_{\!\! A}$ into $N$ row block sub-matrices that have an equal number of rows, and denote their $k$-th row block sub-matrix respectively by $S_{\!\! H,k}$ and $S_{\!\! A,k}$. For each $k=1,2,\cdots,N$, partition the matrix $S_{\!\! H,k}$ further into $2n_{u}$ row block sub-matrices that have an equal number of rows, and denote their $l$-th row block sub-matrix by $S_{\!\! H,k}(l)$. These partitions are always feasible from the definitions of the matrices $S_{\!\! H}$ and $S_{\!\! A}$. On the other hand, to simplify expressions, dependence is omitted for these matrices on both the frequency samples $\omega_{i}|_{i=1}^{N}$ and the parameter vector $\theta^{[0]}$.

With these matrices, define for each $k=1,2,\cdots,N$, further a real matrix $S_{\!\! k}(\omega_{i}|_{i=1}^{N},\theta^{[0]})$ and a complex matrix $\widetilde{S}_{\!\! H,k}(j\omega_{i}|_{i=1}^{N},\theta^{[0]})$ respectively as follows,
\begin{eqnarray*}
& & \hspace*{-1.3cm}
S_{\!\! k}(\omega_{i}|_{i=1}^{N},\theta^{[0]})
=
V_{\Psi}\Sigma_{\Psi}^{-1}U_{\Psi,1}^{T} \left\{ \left[ I \!\otimes\! \overline{\Pi}_{r}(\omega_{i},\theta^{[0]}) \right] S_{\!\! A,k}  + \right. \\
& & \hspace*{4.8cm} \left.  Q_{r}(\omega_{i},\theta^{[0]})S_{\!\! H,k}
 \right\}  \\
& & \hspace*{-1.3cm}
\widetilde{S}_{\!\! H,k}(j\omega_{i}|_{i=1}^{N},\theta^{[0]})\! = \!
\left[V_{zu,1}(j\omega_{k}) \otimes U_{yv,1}(j\omega_{k}) \right]\times \\
& & \hspace*{2.25cm}
\left[ \begin{array}{c}
S_{\!\! H,k}(1)  +  j S_{\!\! H,k}(2) \\
S_{\!\! H,k}(3) + j S_{\!\! H,k}(4) \\
\vdots  \\
S_{\!\! H,k}(2n_{u}-1) + j S_{\!\! H,k}(2n_{u})
\end{array}
\right]
\end{eqnarray*}

On the basis of these symbols, the following parametrization is obtained for all the values of the parameter vector $\theta$ with the corresponding descriptor system ${\rm\bf\Sigma}(\theta)$ having a frequency response slightly different from that of the descriptor system ${\rm\bf\Sigma}(\theta^{[0]})$, measured by the Frobenius norm of a matrix.

\begin{theorem}\label{theo:5}
Assume that $\theta^{[0]}$ is an element of the set ${\rm\bf\Theta}$, and $\omega_{i}|_{i=1}^{N}$ are some distinct frequencies. For a positive number $\varepsilon$, define a parameter set
$\overline{\rm\bf\Theta}_{F}(\varepsilon,\omega_{i}|_{i=1}^{N},\theta^{[0]})$ as
\begin{displaymath}
\overline{\rm\bf\Theta}_{F}(\varepsilon,\omega_{i}|_{i=1}^{N},\theta^{[0]})
\!=\! \left\{ \theta \;\left|\;   \sum_{i=1}^{N}
\left|\left| H(j\omega_{i},\theta) \!-\! H(j\omega_{i},\theta^{[0]})\right|\right|_{F}^{2} \!\leq\! \varepsilon \right.\!\right\}
\end{displaymath}
Assume further that the TFM $G_{zu}(\lambda)$ is of FNRR, and the parameter vector value $\theta^{[0]} \in {\rm\bf\Theta}$ is globally identifiable from the frequency responses of the descriptor system $\rm\bf\Sigma$ at the frequencies $\omega_{i}|_{i=1}^{N}$. Then the set $\overline{\rm\bf\Theta}_{F}(\varepsilon,\omega_{i}|_{i=1}^{N},\theta^{[0]})$ can be equivalently rewritten as
\begin{eqnarray*}
& & \hspace*{-1.2cm} \overline{\rm\bf\Theta}_{F}(\varepsilon,\omega_{i}|_{i=1}^{N},\theta^{[0]})
\\
& &\hspace*{-1.4cm}=\! \left\{ \theta \left| \! \begin{array}{l} \theta = \theta^{[0]} +
S_{\!\! k}(\omega_{i}|_{i=1}^{N},\theta^{[0]})\:\xi + O\left( ||\xi||_{2}^{2} \right),\;\; \xi \in {\cal R}^{n_{s}},\\
\hspace*{0.65cm}{\displaystyle
 \xi^{T}\left( \sum_{l=1}^{N}
\widetilde{S}_{\!\! H,l}^{H}(j\omega_{i}|_{i=1}^{N},\theta^{[0]})
\widetilde{S}_{\!\! H,l}(j\omega_{i}|_{i=1}^{N},\theta^{[0]})
\right) \xi
 \!\leq\! \varepsilon }
\end{array}\right.\!\!\!\!\right\}
\end{eqnarray*}
provided that the positive number $\varepsilon$ is small enough.
\end{theorem}

Note that for every positive $\varepsilon$, the set
\begin{displaymath}
\left\{ \xi \left| \; \xi \in {\cal R}^{n_{s}},{\displaystyle
 \xi^{T}\left( \sum_{l=1}^{N}
\widetilde{S}_{\!\! H,l}^{H}(j\omega_{i}|_{i=1}^{N},\theta^{[0]})
\widetilde{S}_{\!\! H,l}(j\omega_{i}|_{i=1}^{N},\theta^{[0]})
\right) \xi
 \!\leq\! \varepsilon }
\right.\!\right\}
\end{displaymath}
denote it by $\widehat{\rm\bf\Theta}_{F}(\varepsilon,\omega_{i}|_{i=1}^{N},\theta^{[0]})$, is an ellipsoid. This is a quite appealing property that enables an explicit expression for the approximation of both the absolute and the relative sloppiness metrics under a fairly weak condition.

More precisely, we have the following results.

\begin{theorem}\label{theo:6}
Assume that the TFM $G_{zu}(\lambda)$ is of FNRR, and $\theta^{[0]}\in {\rm\bf\Theta}$ is globally identifiable from the frequency responses of the descriptor system $\rm\bf\Sigma$ at the distinct frequencies $\omega_{i}|_{i=1}^{N}$. For each $k \in \{1,2,\cdots,N\}$, let $\mu^{[i]}$ with $i=1,2,\cdots,n_{s}$, be some real scalars satisfying
$\mu^{[1]} \geq \mu^{[2]} \geq \cdots \geq \mu^{[n_{s}]} \geq 0$ and
\begin{eqnarray}
& & \hspace*{-1.2cm}{\rm\bf det}\left[ \mu^{[i]} \left( \sum_{i=1}^{N}
\left(\widetilde{S}_{\!\! H,k,r}^{T}(\omega_{i}|_{i=1}^{N},\theta^{[0]})
\widetilde{S}_{\!\! H,k,r}(\omega_{i}|_{i=1}^{N},\theta^{[0]}) + \right. \right.\right. \nonumber \\
& & \hspace*{0.4cm} \left.\left.
\widetilde{S}_{\!\! H,k,j}^{T}(\omega_{i}|_{i=1}^{N},\theta^{[0]})
\widetilde{S}_{\!\! H,k,j}(\omega_{i}|_{i=1}^{N},\theta^{[0]}) \right)\right) - \nonumber \\
& & \hspace*{2.2cm} \left.
S^{T}_{\!\! k}(\omega_{i}|_{i=1}^{N},\theta^{[0]})
S_{\!\! k}(\omega_{i}|_{i=1}^{N},\theta^{[0]})\right]  =0
\label{eqn:theo6}
\end{eqnarray}
in which $\widetilde{S}_{\!\! H,k,r}(\omega_{i}|_{i=1}^{N},\theta^{[0]})$ and $\widetilde{S}_{\!\! H,k,j}(\omega_{i}|_{i=1}^{N},\theta^{[0]})$ stand respectively for the real and imaginary parts of the complex matrix $\widetilde{S}_{\!\! H,k}(j\omega_{i}|_{i=1}^{N},\theta^{[0]})$. Then the absolute and relative sloppiness metrics
${\rm\bf Sm}^{[a]}_{F,2}(\varepsilon,\omega_{i}|_{i=1}^{N},\theta^{[0]})$ and ${\rm\bf Sm}^{[r,k]}_{F,2}(\varepsilon,\omega_{i}|_{i=1}^{N},\theta^{[0]})$ with $k=1,2,\cdots,n_{s}$, of the descriptor system at the parameter value $\theta^{[0]}$ can be given respectively by the following expressions,
\begin{eqnarray*}
& &
{\rm\bf Sm}^{[a]}_{F,2}(\varepsilon,\omega_{i}|_{i=1}^{N},\theta^{[0]})
= \sqrt{\mu^{[1]}}   \\
& &
{\rm\bf Sm}^{[r,k]}_{F,2}(\varepsilon,\omega_{i}|_{i=1}^{N},\theta^{[0]})
= \sqrt{\frac{\mu^{[k]}}{\mu^{[k+1]}}}
\end{eqnarray*}
\end{theorem}

Note that the product of the transpose of a real matrix and itself is at least positive semi-definite. It is obvious that each $\mu^{[i]}$ satisfying Eq.(\ref{eqn:theo6}) is real and not negative. Hence, the numbers $\mu^{[1]}$, $\mu^{[2]}$, $\cdots$, $\mu^{[n_{s}]}$ of the above theorem are well defined.

On the other hand, it is also extensively adopted in system analysis and synthesis that a (weighted) frequency response estimation error at a prescribed frequency is bounded in its maximum singular value, which is closely related to the ${\cal H}_{\infty}$-norm of a TFM that is widely utilized in robust control and estimation theories \cite{zdg1996,zyl2018}. With the  matrices defined for the situation in which the Frobenius norm of (weighted) frequency response deviations is bounded, define further a set ${\rm\bf \Delta}_{k}(\varepsilon,\omega_{i}|_{i=1}^{N},\theta^{[0]})$ for each $k=1,2,\cdots,N$, and for each $\varepsilon \geq 0$, as follows,
\begin{eqnarray*}
& & \hspace*{-1.3cm}
{\rm\bf \Delta}_{k}(\varepsilon,\omega_{i}|_{i=1}^{N},\theta^{[0]})
= \left\{ \xi \;\left| \; \xi \in {\cal R}^{n_{s}},\;\;
\bar{\sigma}\left(U_{yv,1}(j\omega_{i}) \left[
(S_{\!\! H,k}(1) \; +  \right. \right.\right.\right. \\
& & \hspace*{1.2cm}
j S_{\!\! H,k}(2)) \:\xi \;\; \;\; (S_{\!\! H,k}(3) + j S_{\!\! H,k}(4)) \:\xi \;\;\;\;
\cdots \;\;\;\;  \\
& & \hspace*{0.5cm}
\left.
\left.\left.\left. (S_{\!\! H,k}(2n_{u}-1) + j S_{\!\! H,k}(2n_{u})) \:\xi
\right] V_{zu,1}^{T}(j\omega_{i})\right) \leq \varepsilon \right.\right\}
\end{eqnarray*}

By the same token as the arguments in the proof of Theorem \ref{theo:5}, the following results can be obtained. The details are omitted due to their obviousness.

\begin{corollary}\label{coro:1}
Let $\theta^{[0]}$ be an element of the set ${\rm\bf\Theta}$, and   $\omega_{i}|_{i=1}^{N}$ is a set of distinct real numbers. For any positive number $\varepsilon$, define a parameter set
$\overline{\rm\bf\Theta}_{E}(\varepsilon,\omega_{i}|_{i=1}^{N},\theta^{[0]})$ as
\begin{displaymath}
\overline{\rm\bf\Theta}_{E}(\varepsilon,\omega_{i}|_{i=1}^{N},\theta^{[0]})
\!=\! \left\{ \theta \;\left|  \begin{array}{l}
\bar{\sigma}\left(H(j\omega_{i},\theta) - H(j\omega_{i},\theta^{[0]})\right) \leq \varepsilon \\ \hspace*{0.0cm}
 {\rm for \;\; each}\;\; i \in \left\{ 1,2,\cdots,N \right\}\end{array} \right.\right\}
\end{displaymath}
Then when the parameter vector $\theta^{[0]} \in {\rm\bf\Theta}$ is globally identifiable from the frequency responses of the descriptor system $\rm\bf\Sigma$ at the frequencies $\omega_{i}|_{i=1}^{N}$, the set $\overline{\rm\bf\Theta}(\varepsilon,\omega_{i}|_{i=1}^{N},\theta^{[0]})$ can be expressed equivalently as
\begin{equation}
{\overline{\rm\bf\Theta}}_{E}(\varepsilon,\omega_{i}|_{i=1}^{N},\theta^{[0]})
\!=\! \left\{ \theta \;\left| \! \begin{array}{l} \theta = \theta^{[0]} +
S_{\!\! k}(\omega_{i}|_{i=1}^{N},\theta^{[0]})\:\xi + O\left( ||\xi||_{2}^{2} \right),\\
\hspace*{0.65cm}{\displaystyle \xi \in \bigcap_{l=1}^{N} {\rm\bf \Delta}_{l}(\varepsilon,\omega_{i}|_{i=1}^{N},\theta^{[0]})}
\end{array}\right.\!\!\!\!\right\}
\label{eqn:coro1}
\end{equation}
for each $k \in \left\{ 1,2,\cdots,N\right\}$,
provided that the positive number $\varepsilon$ is small enough.
\end{corollary}

From the definition of the set ${\rm\bf \Delta}_{k}(\varepsilon,\omega_{i}|_{i=1}^{N},\theta^{[0]})$, it is not very difficult to understand that it is convex for each $k=1,2,\cdots,N$. Hence, their union is also convex. This means that the set ${\overline{\rm\bf\Theta}}_{E}(\varepsilon,\omega_{i}|_{i=1}^{N},\theta^{[0]})$ of Eq.(\ref{eqn:coro1}) is approximately convex when $\varepsilon$ is sufficiently small, noting that every element of the set ${\rm\bf \Delta}_{k}(\varepsilon,\omega_{i}|_{i=1}^{N},\theta^{[0]})$ with $k \in \{1,2,\cdots,N\}$, has an Euclidean norm much less than $1$ under this situation.

Based on these observations, it is possible to clarify whether or not there are some sloppy directions for the parameter vector value $\theta^{[0]}$. In particular, for each $k=1,2,\cdots,N$,
define a set $\overline{{\rm\bf \Delta}}_{k}(\varepsilon,\omega_{i}|_{i=1}^{N},\theta^{[0]})$ as
\begin{eqnarray*}
& & \hspace*{-1.00cm} {\rm\bf \overline{\Delta}}_{k}(\varepsilon,\omega_{i}|_{i=1}^{N},\theta^{[0]}) = \left\{\bigcap_{l=1}^{N} {\rm\bf \Delta}_{l}(\varepsilon,\omega_{i}|_{i=1}^{N},\theta^{[0]})\right\}
\bigcap \\
& & \hspace*{-0.5cm}
\left\{ \xi \left|\begin{array}{l}
\xi \!=\! S_{\!\! k}^{\dag}(\omega_{i}|_{i=1}^{N},\theta^{[0]})(\theta \!-\! \theta^{[0]}) + \left[I - S_{\!\! k}^{\dag}(\omega_{i}|_{i=1}^{N},\theta^{[0]})\times \right. \\
\hspace*{4.6cm} \left. S_{\!\! k}(\omega_{i}|_{i=1}^{N},\theta^{[0]})\right]\eta  \\
\hspace*{0.6cm} S_{\!\! k,l}^{\perp}(\omega_{i}|_{i=1}^{N},\theta^{[0]})(\theta - \theta^{[0]}) = 0, \;\;
\theta \in {\rm\bf\Theta},\;\; \eta \in {\cal R}^{n_{s}}
\end{array}\right.\!\!\right\}
\end{eqnarray*}
Then the search for a sloppy direction of the parameter vector value $\theta^{[0]}$ can in principle be implemented as follows.

\begin{itemize}
\item Initialize the vector $\xi^{[0]}$  as $\xi^{[0]} = 0_{n_{s}}$.
\item For each $i=1,2,\cdots,n_{s}$, compute a vector $\xi^{[i]}$ and a quantity $w^{[i]}$ recursively as follows
\begin{eqnarray*}
& & \hspace*{-1.4cm} \xi^{[i]} = \arg \; \max_{\xi}
\xi^{T}
S_{\!\! k}^{T}(\omega_{i}|_{i=1}^{N},\theta^{[0]}) S_{\!\! k}(\omega_{i}|_{i=1}^{N},\theta^{[0]})\:\xi  \\
& & \hspace*{-0.4cm} {\rm s.\; t.\;}  \xi \in
\overline{{\rm\bf \Delta}}_{k}(\varepsilon,\omega_{i}|_{i=1}^{N},\theta^{[0]})\backslash {\rm\bf span}\{\xi^{[0]},\;\xi^{[1]},\;\cdots,\;\xi^{[i-1]}\} \\
& & \hspace*{-1.4cm}
w^{[i]} =  \xi^{[i]T}
S_{\!\! k}^{T}(\omega_{i}|_{i=1}^{N},\theta^{[0]}) S_{\!\! k}(\omega_{i}|_{i=1}^{N},\theta^{[0]})\:\xi^{[i]}
\end{eqnarray*}
\end{itemize}

In this procedure, the integer $k$ can be selected to be any element of the set  $\{1,2,3,\cdots,N\}$, and the computed $w^{[1]}$, $w^{[2]}$, $\cdots$, $w^{[n_{s}]}$  in principle do not depend on its choice. On the other hand, the obtained $w^{[1]}|_{i=1}^{[n_{s}]}$ approximately reveal information of the distances from the parameter vector value $\theta^{[0]}$ to the boundary of the set $\overline{\rm\bf\Theta}_{E}(\varepsilon,\omega_{i}|_{i=1}^{N},\theta^{[0]}) \bigcap {\rm\bf\Theta}$ in each extreme direction, from which existence of sloppy directions can be detected. In addition, the absolute and relative sloppiness metrics
${\rm\bf Sm}^{[a]}_{E,2}(\varepsilon,\omega_{i}|_{i=1}^{N},\theta^{[0]})$ and ${\rm\bf Sm}^{[r,k]}_{E,2}(\varepsilon,\omega_{i}|_{i=1}^{N},\theta^{[0]})$ of the descriptor system $\rm\bf\Sigma$ at the parameter value $\theta^{[0]}$ can be directly shown to have respectively the following expressions,
\begin{eqnarray*}
& & {\rm\bf Sm}^{[a]}_{E}(\varepsilon,\omega_{i}|_{i=1}^{N},\theta^{[0]})
= \frac{\sqrt{w^{[1]}}}{\varepsilon} + O(\varepsilon^{2})\\
& &
{\rm\bf Sm}^{[r]}_{E}(\varepsilon,\omega_{i}|_{i=1}^{N},\theta^{[0]})
= \sqrt{\frac{w^{[1]}}{w^{[n_{s}]}}} + O(\varepsilon^{2})
\end{eqnarray*}

However, it is worthwhile to point out that the optimizations involved in the above procedure include a problem of maximizing a convex function over a convex set, which is well known to be NP-hard in general \cite{zyl2018}. A possible solution is to replace the set $\overline{{\rm\bf\Delta}}_{k}(\varepsilon,\omega_{i}|_{i=1}^{N},\theta^{[0]})$ with an ellipsoid which include this set as a subset and has the minimum volume in searching the maximum distance from its boundary to the parameter vector value $\theta^{[0]}$ , while replace this set with an ellipsoid which is contained in this set as a subset and has the maximum volume in searching the minimum distance from its boundary to the parameter vector value $\theta^{[0]}$. When these replacements are available, an explicit expression can be obtained respectively for an approximation to the absolute sloppiness metric and the relative sloppiness metric, which are similar to that when the Frobenius norm is adopted to measure frequency response deviations.

\section{Concluding Remarks}

This paper investigates identifiability and sloppiness for the parameters of a descriptor system when they are required to be estimated from its frequency responses. Under the condition that system matrices depend on its parameters through an LFT and an associated TFM is of FNRR, a necessary and sufficient condition is derived for global parameter identifiability at a particular value with a set of finitely many frequency responses. This condition is based on the rank of a matrix, and can be verified recursively. The latter is computationally quite attractive in analysis and synthesis of a large scale NDS. An algorithm is suggested to find a set of frequencies with which the frequency responses of the system are capable to uniquely determine its parameters. Two metrics are proposed  respectively for measuring absolute and relative sloppiness of the parameter vector at a prescribed value. An ellipsoid approximation is given for the set consisting of all the parameter values with which the frequency response of the corresponding descriptor system deviates within a prescribed distance, from that corresponding to a globally identifiable parameter vector value. Explicit formulas are also obtained for the suggested absolute and relative sloppiness metrics.

\renewcommand{\labelenumi}{\rm\bf A\arabic{enumi})}

\renewcommand{\theequation}{a\arabic{equation}}
\setcounter{equation}{0}

\small
\section*{\small \hspace*{-0.2cm} Appendix: Proof of Some Technical Results}

\noindent\textbf{Proof of Lemma \ref{lemma:3}:} Let $D_{l}^{-1}(\lambda, \theta)N_{l}(\lambda, \theta)$ and $N_{r}(\lambda, \theta)D^{-1}_{r}(\lambda, \theta)$ denote respectively the left an right coprime factorizations of the TFM $H(\lambda, \theta)$ of the descriptor system $\rm\bf\Sigma$, in which $D_{l}(\lambda, \theta)$, $D_{r}(\lambda, \theta)$, $N_{l}(\lambda, \theta)$ and $N_{r}(\lambda, \theta)$ are MVPs of degree $M$ with $M$ being a finite positive integer.

Assume that the descriptor system $\rm\bf\Sigma$ is of continuous time. Moreover, assume that there are two parameter vectors $\theta \in {\rm\bf\Theta}$ and $\widetilde{\theta} \in {\rm\bf\Theta}$ that satisfy the following equality
\begin{equation}
H(j\omega_{i},\theta) =  H(j\omega_{i},\widetilde{\theta})
\end{equation}
at $N$ distinct frequency samples $\omega_{i}$, $i=1,2,\cdots,N$. Then according to the left and right coprime factorizations of the associated TFM, we have that for each $i=1,2,\cdots,N$,
\begin{equation}
D_{l}(j\omega_{i},\widetilde{\theta})N_{r}(j\omega_{i},{\theta}) =  N_{l}(j\omega_{i},\widetilde{\theta})D_{r}(j\omega_{i},{\theta})
\label{eqn:cop-fac}
\end{equation}

On the other hand, as all the involved MVPs have a finite degree $M$, it is obvious that there exist real valued matrices $A_{i}(\theta,\widetilde{\theta})$ and $B_{i}(\theta,\widetilde{\theta})$ with  $i=0,1,\cdots,2M$, such that
\begin{equation}
D_{l}(\lambda,\widetilde{\theta})N_{r}(\lambda,{\theta}) = \sum_{i=0}^{2M} A_{i}(\theta,\widetilde{\theta})\lambda^{i},\hspace{0.15cm} N_{l}(\lambda,\widetilde{\theta})D_{r}(\lambda,{\theta}) = \sum_{i=0}^{2M} B_{i}(\theta,\widetilde{\theta})\lambda^{i}
\label{eqn:cop-fac-2}
\end{equation}
This implies that Eq.(\ref{eqn:cop-fac}) can be equivalently rewritten as
\begin{eqnarray}
& &\hspace*{-1cm} \left[ A_{0}(\theta,\widetilde{\theta})\;\; A_{1}(\theta,\widetilde{\theta})\;\; \cdots\;\; A_{2M}(\theta,\widetilde{\theta}) \right]\left( W \left(\omega_{i}|_{i=1}^{N}\right) \otimes I \right) \nonumber\\
& & \hspace*{-1.2cm}= \left[ B_{0}(\theta,\widetilde{\theta})\;\; B_{1}(\theta,\widetilde{\theta})\;\; \cdots\;\; B_{2M}(\theta,\widetilde{\theta}) \right]\left( W \left(\omega_{i}|_{i=1}^{N}\right)\otimes I \right)
\label{eqn:cop-fac-3}
\end{eqnarray}
in which
\begin{displaymath}
W \left(\omega_{i}|_{i=1}^{N}\right) = \left[\begin{array}{cccc}
1 & 1 & \cdots & 1 \\
j\omega_{1} & j\omega_{2} & \cdots & j\omega_{N} \\
\vdots & \vdots & \ddots & \vdots \\
(j\omega_{1})^{2M} & (j\omega_{2})^{2M} & \cdots & (j\omega_{N})^{2M} \end{array}\right]\
\end{displaymath}

It is obvious from matrix analyzes \cite{hj1991} that when $N\geq 2M + 1$ and $\omega_{i}|_{i=1}^{N}$ are different, the matrix $W \left(\omega_{i}|_{i=1}^{N}\right)$ is always of FRR, which further implies that the Kronecker product $W \left(\omega_{i}|_{i=1}^{N}\right)\otimes I$ is of FRR. As a matter of fact, when $N = 2M + 1$, the aforementioned matrix $W \left(\omega_{i}|_{i=1}^{N}\right)$ is actually an Vandermonde matrix whose determinant is not equal to zero, provided that the frequencies $\omega_{1}$, $\omega_{2}$, $\cdots$, $\omega_{2M + 1}$ are distinct from each other. It can therefore be declared from Eq.(\ref{eqn:cop-fac-3}) that
\begin{eqnarray}
& & \left[ A_{0}(\theta,\widetilde{\theta})\;\; A_{1}(\theta,\widetilde{\theta})\;\; \cdots\;\; A_{2M}(\theta,\widetilde{\theta}) \right]\nonumber\\
&=& \left[ B_{0}(\theta,\widetilde{\theta})\;\; B_{1}(\theta,\widetilde{\theta})\;\; \cdots\;\; B_{2M}(\theta,\widetilde{\theta}) \right]
\end{eqnarray}

On the basis of this equality and Eq.(\ref{eqn:cop-fac-2}), we have that
\begin{equation}
D_{l}(\lambda,\widetilde{\theta})N_{r}(\lambda,{\theta}) \equiv N_{l}(\lambda,\widetilde{\theta})D_{r}(\lambda,{\theta})
\end{equation}
which is equivalent to that
\begin{equation}
H(\lambda,\widetilde{\theta})= H(\lambda,{\theta})
\end{equation}
The proof can now be completed through an application of Lemma \ref{lemma:2}.
\hspace{\fill}$\Diamond$

\noindent\textbf{Proof of Lemma \ref{lemma:4}:} Denote the real and imaginary parts of the matrices $V_{\Pi}$ and $U_{\Pi,1}$ respectively by $V_{\Pi,r}$ and $V_{\Pi,j}$, $U_{\Pi,1r}$ and $U_{\Pi,1j}$. Then from the SVD of the matrix $\Pi$, we have that
\begin{eqnarray}
\hspace*{-0.6cm} \Pi_{r} + j\Pi_{j} &=& (U_{\Pi,1r} + jU_{\Pi,1j}) \Sigma (V^{T}_{\Pi,r} -j V^{T}_{\Pi,j} ) \nonumber \\
& &\hspace*{-1.8cm}= (U_{\Pi,1r}\Sigma V^{T}_{\Pi,r}+ j U_{\Pi,1j}\Sigma V^{T}_{\Pi,j}) + j (-U_{\Pi,1r}\Sigma V^{T}_{\Pi,j} + U_{\Pi,1j}\Sigma V^{T}_{\Pi,r} )
\end{eqnarray}
We therefore have that
\begin{eqnarray}
& & \hspace{-1.4cm} \left[\!\!\begin{array}{rr} \Pi_{r} & -\Pi_{j} \\ \Pi_{j} & \Pi_{r} \end{array} \!\!\right] =
\left[\!\!\begin{array}{rr} U_{\Pi,1r} & -U_{\Pi,1j} \\ U_{\Pi,1j} & U_{\Pi,1r} \end{array} \!\!\right]
\left[\!\!\begin{array}{rr} \Sigma & 0 \\ 0 & \Sigma \end{array} \!\!\right]
\left[\!\!\begin{array}{rr} V_{\Pi,r} & -V_{\Pi,j} \\ V_{\Pi,j} & V_{\Pi,r} \end{array} \!\!\right]^{T}   \\
& & \hspace{-1.4cm} \left[\!\!\begin{array}{rr} \Pi_{r} & -\Pi_{j} \end{array}\!\!\right] =
\left[\!\!\begin{array}{rr} U_{\Pi,1r} & -U_{\Pi,1j} \end{array} \!\!\right]
\left[\!\!\begin{array}{rr} \Sigma & 0 \\ 0 & \Sigma \end{array} \!\!\right]
\left[\!\!\begin{array}{rr} V_{\Pi,r} & -V_{\Pi,j} \\ V_{\Pi,j} & V_{\Pi,r} \end{array} \!\!\right]^{T}
\end{eqnarray}

On the other hand, from the fact that both the matrix $\left[ U_{\Pi,1}\;\; U_{\Pi,2} \right]$ and the matrix $V_{\Pi}$ are unitary, it can be directly shown that
\begin{eqnarray}
& & \hspace{-1.4cm}
\left[\!\!\begin{array}{rr} U_{\Pi,1r} & -U_{\Pi,1j} \\ U_{\Pi,1j} & U_{\Pi,1r} \end{array} \!\!\right]\left[\!\!\begin{array}{rr} U_{\Pi,1r} & -U_{\Pi,1j} \\ U_{\Pi,1j} & U_{\Pi,1r} \end{array} \!\!\right]^{T}= I \\
& & \hspace{-1.4cm}
\left[\!\!\begin{array}{rr} V_{\Pi,r} & -V_{\Pi,j} \\ V_{\Pi,j} & V_{\Pi,r} \end{array} \!\!\right]\left[\!\!\begin{array}{rr} V_{\Pi,r} & -V_{\Pi,j} \\ V_{\Pi,j} & V_{\Pi,r} \end{array} \!\!\right]^{T}=I
\label{eqn:svd-1}\\
& & \hspace{-1.4cm}
U_{\Pi,1r} U^{T}_{\Pi,1r} + U_{\Pi,1j}U^{T}_{\Pi,1j} +
U_{\Pi,2r} U^{T}_{\Pi,2r} + U_{\Pi,2j}U^{T}_{\Pi,2j} = I
\end{eqnarray}
As the unitary matrix $V_{\Pi}$ is square, Eq.(\ref{eqn:svd-1}) implies further that the matrix $\left[\!\!\begin{array}{rr} V_{\Pi,r} & -V_{\Pi,j} \\ V_{\Pi,j} & V_{\Pi,r} \end{array} \!\!\right]$ is invertible.

Therefore
\begin{eqnarray}
& &\hspace*{-0.5cm} I- \left[\!\!\begin{array}{cc} \Pi_{r} & -\Pi_{j} \end{array} \!\!\right]
\left(\left[\!\!\begin{array}{rr} \Pi_{r} & -\Pi_{j} \\ \Pi_{j} & \Pi_{r} \end{array} \!\!\right]^{T}\left[\!\!\begin{array}{rr} \Pi_{r} & -\Pi_{j} \\ \Pi_{j} & \Pi_{r} \end{array} \!\!\right]\right)^{-1}
\!\!\left[\!\!\begin{array}{cc} \Pi_{r} & -\Pi_{j} \end{array} \!\! \right]^{T} \nonumber \\
& &\hspace*{-0.9cm} =
I-\left(\left[\!\!\begin{array}{rr} U_{\Pi,1r} & -U_{\Pi,1j} \end{array} \!\!\right]
\left[\!\!\begin{array}{rr} \Sigma & 0 \\ 0 & \Sigma \end{array} \!\!\right]
\left[\!\!\begin{array}{rr} V_{\Pi,r} & -V_{\Pi,j} \\ V_{\Pi,j} & V_{\Pi,r} \end{array} \!\!\right]\right)\times \nonumber \\
& &\hspace*{0.5cm}
\left\{\left(\left[\!\!\begin{array}{rr} U_{\Pi,1r} & -U_{\Pi,1j} \\ U_{\Pi,1j} & U_{\Pi,1r} \end{array} \!\!\right]
\left[\!\!\begin{array}{rr} \Sigma & 0 \\ 0 & \Sigma \end{array} \!\!\right]
\left[\!\!\begin{array}{rr} V_{\Pi,r} & -V_{\Pi,j} \\ V_{\Pi,j} & V_{\Pi,r} \end{array} \!\!\right]\right)^{T}\times\right. \nonumber\\
& &\hspace*{0.5cm}
\left.\left(\left[\!\!\begin{array}{rr} U_{\Pi,1r} & -U_{\Pi,1j} \\ U_{\Pi,1j} & U_{\Pi,1r} \end{array} \!\!\right]
\left[\!\!\begin{array}{rr} \Sigma & 0 \\ 0 & \Sigma \end{array} \!\!\right]
\left[\!\!\begin{array}{rr} V_{\Pi,r} & -V_{\Pi,j} \\ V_{\Pi,j} & V_{\Pi,r} \end{array} \!\!\right]\right)^{T}\right\}^{-1}\times \nonumber\\
& & \hspace*{0.5cm}
\left(\left[\!\!\begin{array}{rr} U_{\Pi,1r} & -U_{\Pi,1j} \end{array} \!\!\right]
\left[\!\!\begin{array}{rr} \Sigma & 0 \\ 0 & \Sigma \end{array} \!\!\right]
\left[\!\!\begin{array}{rr} V_{\Pi,r} & -V_{\Pi,j} \\ V_{\Pi,j} & V_{\Pi,r} \end{array} \!\!\right]\right)^{T} \nonumber \\
& &\hspace*{-0.9cm} = I - \left[\!\!\begin{array}{cc} U_{\Pi,1r} & -U_{\Pi,1j} \end{array} \!\!\right]
\left[\!\!\begin{array}{cc} U_{\Pi,1r} & -U_{\Pi,1j} \end{array} \!\!\right]^{T}
\nonumber \\
& &\hspace*{-0.9cm} = \left[\!\!\begin{array}{cc} U_{\Pi,2r} & U_{\Pi,2j} \end{array} \!\!\right]
\left[\!\!\begin{array}{cc} U_{\Pi,2r} & U_{\Pi,2j} \end{array} \!\!\right]^{T}
\end{eqnarray}

This completes the proof.  \hspace{\fill}$\Diamond$

\noindent\textbf{Proof of Theorem \ref{theo:1}:} Assume that there is a $\theta^{[\star]} \in {\rm\bf\Theta}$ which is different from $\theta^{[0]}$ and satisfies $H(\lambda,\theta^{[\star]})-H(\theta,\Phi_0) \equiv 0$. Moreover, assume that
$\theta^{[0]} = col\left\{\theta^{[0]}_{k}|_{k=1}^{q}\right\}$ and $\theta^{[\star]} = col\left\{\theta^{[\star]}_{k}|_{k=1}^{q}\right\}$. Then according to the definition of the TFM $H(\lambda,\theta)$ given in Eq.(\ref{eqn:tfm}), we have that
\begin{eqnarray}
& &\hspace*{-0.6cm}H(\lambda,\theta^{[\star]})-H(\lambda,\theta^{[0]}) \nonumber \\
& & \hspace*{-1cm} = G_{yv}(\lambda)\left\{P(\theta^{[\star]})[I_{m_z}-G_{zv}(\lambda)P(\theta^{[\star]})]^{-1}\!- \right.  \nonumber \\
& & \hspace*{1.5cm}\left. [I_{m_v}\!-P(\theta^{[0]})G_{zv}(\lambda)]^{-1}P(\theta^{[0]})\right\}G_{zu}(\lambda)
\nonumber \\
& &\hspace*{-1cm} = G_{yv}(\lambda)[I_{m_v}-P(\theta^{[0]})G_{zv}(\lambda)]^{-1}\left[\sum_{k=1}^{q}
(\theta^{[0]}_{k}-\theta^{[\star]}_{k})P_{k}\right] \times \nonumber\\
& & \hspace*{2.8cm} [I_{m_z}-G_{zv}(\lambda)P(\theta^{[\star]})]^{-1}G_{zu}(\lambda)  \nonumber\\
& & \hspace*{-1cm}\equiv 0
\label{a.11}
\end{eqnarray}
According to Lemma $\ref{lemma:2}$, the structure of the descriptor system $\rm\bf\Sigma$ is not globally identifiable at this specific parameter vector $\theta^{[0]} \in {\rm\bf\Theta}$, if and only if there is a parameter vector $\theta^{[\star]} \in {\rm\bf\Theta}$ that is different from the parameter vector $\theta^{[0]}$ and satisfies the above equation.

Note that
\begin{eqnarray}
vec\left(\sum_{k=1}^{q}
(\theta^{[0]}_{k}-\theta^{[\star]}_{k})P_{k}\right)
&=& \sum_{k=1}^{q}
(\theta^{[0]}_{k}-\theta^{[\star]}_{k})vec(P_{k}) \nonumber \\
&=& \Psi (\theta^{[0]} - \theta^{[\star]})	
\end{eqnarray}
It is obvious that if the matrix $\Psi$ is not of FCR and the descriptor system ${\rm\bf\Sigma}(\theta^{[0]})$ is regular and well-posed, then there certainly exists a parameter vector $\theta^{[\star]} \in {\rm\bf\Theta}$, which is different from the parameter vector $\theta^{[0]}$, such that the associated descriptor system ${\rm\bf\Sigma}(\theta^{[0]})$ is regular, and $H(\lambda,\theta^{[\star]})-H(\lambda,\theta^{[0]}) \equiv 0$. That is, the descriptor system ${\rm\bf\Sigma}$ is not globally identifiable at this particular parameter vector $\theta^{[0]}$. This further means that to guarantee that the descriptor system ${\rm\bf\Sigma}$ is globally identifiable, it is necessary that the matrix $\Psi$ is of FCR.

Assume now that the matrix $\Psi$ is of FCR and the TFM $G_{zu}(\lambda)$ is of FNRR. Note that the invertibility of the TFM $I_{m_z}-G_{zv}(\lambda)P(\theta^{[\star]})$ is guaranteed by the regularity and well-posedness assumptions of the descriptor system ${\rm\bf\Sigma}(\theta^{[\star]})$. Then it is obvious from Eq.(\ref{a.11}) that $H(\lambda,\theta^{[\star]})-H(\lambda,\theta^{[0]}) \equiv 0$ is equivalent to
\begin{equation}
G_{yv}(\lambda)[I_{m_v}-P(\theta^{[0]})G_{zv}(\lambda)]^{-1}\left[\sum_{k=1}^{q}
(\theta^{[0]}_{k}-\theta^{[\star]}_{k})P_{k}\right] \equiv 0
\label{a.12}
\end{equation}

From the Smith-McMillan form of the TFM $G_{yv}(\lambda)$, the above equation can be further equivalently expressed as
\begin{equation}
U_{yv, 1}(\lambda)\Sigma_{yv}(\lambda)V^{T}_{yv, 1}(\lambda)[I_{m_v}-P(\theta^{[0]})G_{zv}(\lambda)]^{-1}\left[\sum_{k=1}^{q}
(\theta^{[0]}_{k}-\theta^{[\star]}_{k})P_{k}\right] \equiv 0
\label{a.13}
\end{equation}

Recall that the MVP $\left[ U_{yv, 1}(\lambda)\;\;U_{yv, 2}(\lambda) \right]$ is unimodular, while $\Sigma_{yv}(\lambda)$ is diagonal with every diagonal element being a nonzero rational function. It is obvious that both $U_{yv, 1}(\lambda)$ and $\Sigma_{yv}(\lambda)$ are of FNCR. We therefore have that Eq.(\ref{a.13})  is equivalent to
\begin{equation}
V_{yv, 1}(\lambda)[I_{m_v}-P(\theta^{[0]})G_{zv}(\lambda)]^{-1}\left[\sum_{k=1}^{q}
(\theta^{[0]}_{k}-\theta^{[\star]}_{k})P_{k}\right] \equiv 0
\label{a.14}
\end{equation}

On the basis of this equality, arguments similar to the proof of Lemma \ref{lemma:3} show that when that when the TFM $G_{zu}(\lambda)$ is of FNRR, the descriptor system ${\rm\bf\Sigma}$ is globally identifiable at a particular parameter vector $\theta^{[0]}$, if and only if there exists a set of frequency samples, say $\omega_{i}|_{i=1}^{N}$, with finitely many elements, such that for every parameter vector $\theta^{[\star]} \in {\rm\bf\Theta}$, the satisfaction of the following equality
\begin{equation}
V^{T}_{yv,1}(j\omega_{i})[I_{m_v}-P(\theta^{[0]})G_{zv}(j\omega_{i})]^{-1}\left[\sum_{k=1}^{q}
(\theta^{[0]}_{k}-\theta^{[\star]}_{k})P_{k}\right] = 0
\label{eqn:a15}
\end{equation}
at each $i=1,2,\cdots,N$, implies that $\theta^{[\star]} = \theta^{[0]}$.

Note that the MVP $\left[ V_{yv, 1}(\lambda)\;\;V_{yv, 2}(\lambda) \right]$ is unimodular. We therefore have that $\left[ V_{yv, 1}(\lambda)\;\;V_{yv, 2}(\lambda) \right]^{-1}$ is also unimodular that is well defined at each $\lambda \in {\cal C}$. These mean that at every $\omega\in {\cal R}$, the following equality
\begin{equation}
\left[\begin{array}{c} V^{[inv,1]}_{yv}(j\omega) \\ V^{[inv,2]}_{yv}(j\omega) \end{array}\right]
\left[ V_{yv, 1}(j\omega)\;\;V_{yv, 2}(j\omega) \right] = I
\end{equation}
is valid, and the complex matrix $V^{[inv,2]}_{yv}(j\omega)$ is of FRR.

From these relations and the definition of the MVF $\Pi(\lambda,\theta)$ given in Eq.(\ref{eqn:pi}), it is not hard to show that Eq.(\ref{eqn:a15}) is satisfied by a parameter vector $\theta^{[\star]}$, if and only if there exists a complex matrix $A(j\omega_{i})$, such that
\begin{equation}
\left[\sum_{k=1}^{q}
(\theta^{[0]}_{k}-\theta^{[\star]}_{k})P_{k}\right] = \Pi(j\omega_{i},\theta^{[0]})A(j\omega_{i})
\label{a.16}
\end{equation}

Let $A_{r}(\omega_{i})$ and $A_{j}(\omega_{i})$ stand respectively for the real and imaginary parts of the complex matrix $A(j\omega_{i})$. Moreover, denote for brevity the matrices $\left[\Pi_{r}(\omega_{i},\theta^{[0]})\;\; -\Pi_{j}(\omega_{i},\theta^{[0]})\right]$, $\left[\Pi_{j}(\omega_{i},\theta^{[0]})\;\; \Pi_{r}(\omega_{i},\theta^{[0]})\right]$ and $col\left\{A_{r}(\omega_{i})  \;\;  A_{j}(\omega_{i}) \right\}$ respectively by $\overline{\Pi}_{r}(\omega_{i}, \theta^{[0]})$, $\overline{\Pi}_{j}(\omega_{i}, \theta^{[0]})$ and $\overline{A}(\omega_{i})$. Note that all the matrices $P_{k}|_{k=1}^{q}$, as well as all the scalars $\theta^{[0]}_{k}|_{k=1}^{q}$ and $\theta^{[\star]}_{k}|_{k=1}^{q}$, are real. The above equality can be equivalently rewritten as
\begin{eqnarray}
& &\hspace*{-0.20cm} \left[\sum_{k=1}^{q}
(\theta^{[0]}_{k}-\theta^{[\star]}_{k})P_{k}\right] = \overline{\Pi}_{r}(\omega_{i}, \theta^{[0]})\overline{A}(\omega_{i})
\label{eqn:a17}   \\
& & \hspace*{-0.20cm} \overline{\Pi}_{j}(\omega_{i}, \theta^{[0]})\overline{A}(\omega_{i}) = 0
\label{eqn:a18}
\end{eqnarray}

From the definition of the matrix $\Psi$ given in Eq.(\ref{eqn:par-vec-1}) and the vectorization of both sides of Eqs.(\ref{eqn:a17}) and (\ref{eqn:a18}), we further have that
\begin{eqnarray}
& &\hspace*{-0.20cm} \Psi
(\theta^{[0]}-\theta^{[\star]}) = \left(I\otimes\overline{\Pi}_{r}(\omega_{i}, \theta^{[0]})\right){\rm\bf vec}(\overline{A}(\omega_{i}))
\label{eqn:a19}   \\
& & \hspace*{-0.20cm} \left(I\otimes\overline{\Pi}_{j}(\omega_{i}, \theta^{[0]})\right){\rm\bf vec}(\overline{A}(\omega_{i})) = 0
\label{eqn:a20}
\end{eqnarray}

When the matrix $\Psi$ is of FCR, the matrix $V_{\Psi,2}$ of Eq.(\ref{eqn:par-vec-2}) is empty. On the other hand, from the SVD of the matrix $\Psi$ given by Eq.(\ref{eqn:par-vec-2}), we have that the matrix $U_{\Psi,2}$ is of FCR and $U^{T}_{\Psi,2} \Psi =0$. It can be directly declared from Lemma \ref{lemma:1} that, there exists a parameter vector $\theta^{[\star]}$ satisfying Eq.(\ref{eqn:a19}) if and only if
\begin{equation}
U^{T}_{\Psi,2} \left(I\otimes\overline{\Pi}_{r}(\omega_{i}, \theta^{[0]})\right){\rm\bf vec}(\overline{A}(\omega_{i})) = 0
\label{eqn:a21}
\end{equation}
Moreover, for a particular complex matrix $\overline{A}(\omega_{i})$ satisfying this condition, there is an unique $\theta^{[\star]}$ that satisfies Eq.(\ref{eqn:a19}) which can be expressed as follows,
\begin{eqnarray}
\theta^{[\star]} \!\!&=&\!\!\!\! \theta^{[0]} + (\Psi^{T}\Psi)^{-1}\Psi^{T} \left(I\otimes\overline{\Pi}_{r}(\omega_{i}, \theta^{[0]})\right){\rm\bf vec}(\overline{A}(\omega_{i})) \nonumber \\
&=&\!\!\!\! \theta^{[0]} \!+\! V_{\Psi} \Sigma^{-1}_{\Psi}U^{T}_{\Psi,1}   \left(I\otimes\overline{\Pi}_{r}(\omega_{i}, \theta^{[0]})\right)\!{\rm\bf vec}(\overline{A}(\omega_{i}))
\label{eqn:a22}
\end{eqnarray}

Assume now that there exists a frequency sample $\omega_{1}$, such that the matrix $\overline{\Pi}_{j}(\omega_{1}, \theta^{[0]})$ is of FCR. Then according to the properties of Kronecker matrix products, the matrix  $I\otimes\overline{\Pi}_{j}(\omega_{1}, \theta^{[0]})$ is also FCR. It can therefore be declared that at this frequency sample, Eq.(\ref{eqn:a20}) is satisfied only by the vector ${\rm\bf vec}(\overline{A}(\omega_{1})) =0$. On the basis of Eq.(\ref{eqn:a22}), this further means that $\theta^{[\star]} = \theta^{[0]}$.

Now assume that for each $\omega \in {\cal R}$, the matrix $\overline{\Pi}_{j}(\omega, \theta^{[0]})$ is not of FCR. For a given series of frequencies $\omega_{1}$, $\omega_{2}$, $\cdots$, $\omega_{N}$, define a real matrix $\Gamma(\omega_{i}|_{i=1}^{N},\theta^{[0]})$ as follows,
\begin{displaymath}
\Gamma(\omega_{i}|_{i=1}^{N},\theta^{[0]}) =
\left[\begin{array}{c}
\Gamma_{r,1}(\omega_{i}|_{i=1}^{N},\theta^{[0]})  \\
\Gamma_{r,2}(\omega_{i}|_{i=1}^{N},\theta^{[0]})  \\
\Gamma_{j}(\omega_{i}|_{i=1}^{N},\theta^{[0]})
\end{array}\right]
\end{displaymath}
in which the sub-matrices $\Gamma_{r,1}(\omega_{i}|_{i=1}^{N},\theta^{[0]})$, $\Gamma_{r,2}(\omega_{i}|_{i=1}^{N},\theta^{[0]})$ and $\Gamma_{j}(\omega_{i}|_{i=1}^{N},\theta^{[0]})$ have respectively the following definitions,
\begin{eqnarray*}
& & \hspace{-1.0cm}
\Gamma_{r,1}(\omega_{i}|_{i=1}^{N},\theta^{[0]}) =
\left[ \Gamma_{r,11}(\omega_{i}|_{i=1}^{N},\theta^{[0]})\;\;\; \Gamma_{r,12}(\omega_{i}|_{i=1}^{N},\theta^{[0]}) \right]   \\
& & \hspace{-0.5cm}
\Gamma_{r,11}(\omega_{i}|_{i=1}^{N},\theta^{[0]}) =
\left[\begin{array}{c}
U^{T}_{\Psi,1}   \left(I\otimes\overline{\Pi}_{r}(\omega_{1}, \theta^{[0]})\right)  \\
U^{T}_{\Psi,1}   \left(I\otimes\overline{\Pi}_{r}(\omega_{1}, \theta^{[0]})\right)  \\
\vdots  \\
U^{T}_{\Psi,1}   \left(I\otimes\overline{\Pi}_{r}(\omega_{1}, \theta^{[0]})\right)
\end{array}\right] \\
& & \hspace{-0.5cm}
\Gamma_{r,12}(\omega_{i}|_{i=1}^{N},\theta^{[0]}) =
{\rm\bf diag}\left\{
-U^{T}_{\Psi,1}   \left(I\otimes\overline{\Pi}_{r}(\omega_{i}, \theta^{[0]})\right)|_{i=2}^{N} \right\} \\
& & \hspace{-1.0cm}
\Gamma_{r,2}(\omega_{i}|_{i=1}^{N},\theta^{[0]}) = {\rm\bf diag}\left\{
U^{T}_{\Psi,2}   \left(I\otimes\overline{\Pi}_{r}(\omega_{i}, \theta^{[0]})\right)|_{i=1}^{N} \right\}  \\
& & \hspace{-1.0cm}
\Gamma_{j}(\omega_{i}|_{i=1}^{N},\theta^{[0]}) = {\rm\bf diag}\left\{
\left(I\otimes\overline{\Pi}_{j}(\omega_{i}, \theta^{[0]})\right)|_{i=1}^{N} \right\}
\end{eqnarray*}

Then the equalities of Eqs.(\ref{eqn:a20})-(\ref{eqn:a22}) mean that there exists a parameter vector $\theta^{[\star]}$, such that the constraint of Eq.(\ref{eqn:a15}) is simultaneously satisfied at each $\omega_{i}$ with $i=1,2,\cdots,N$, if and only if there exist a series of complex matrices the complex matrix $A(j\omega_{i})$ such that
\begin{equation}
\Gamma(\omega_{i}|_{i=1}^{N},\theta^{[0]}) \;
{\rm\bf col} \left\{ \left.
{\rm\bf vec}(\overline{A}(\omega_{i})) \right|_{i=1}^{N} \right\}
= 0
\label{eqn:a28}
\end{equation}
Moreover, there is only one parameter vector $\theta^{[\star]}$ that simultaneously satisfies these constraints, which is actually the parameter vector $\theta^{[0]}$ itself, if and only if the matrix $\Gamma(\omega_{i}|_{i=1}^{N},\theta^{[0]})$ is of FCR.

Note that a real matrix is of FCR, when and only when the product of its transpose and itself is positive definite. The above arguments mean that the descriptor system $\rm\bf\Sigma$ is globally identifiable at the particular vector $\theta^{[0]}$, if and only if there exist a series of frequency samples $\omega_{i}$, $i=1,2,\cdots,N$, such that
\begin{eqnarray}
& & \hspace*{-1.4cm} \Gamma^{T}(\omega_{i}|_{i=1}^{N},\theta^{[0]}) \Gamma(\omega_{i}|_{i=1}^{N},\theta^{[0]})=
\Gamma^{T}_{r,1}(\omega_{i}|_{i=1}^{N},\theta^{[0]}) \Gamma_{r,1}(\omega_{i}|_{i=1}^{N},\theta^{[0]})
+  \nonumber\\
& & \hspace*{2.4cm}
\Gamma^{T}_{r,2}(\omega_{i}|_{i=1}^{N},\theta^{[0]})  \Gamma_{r,2}(\omega_{i}|_{i=1}^{N},\theta^{[0]})
+ \nonumber\\
& & \hspace*{2.4cm}
\Gamma^{T}_{j}(\omega_{i}|_{i=1}^{N},\theta^{[0]})
\Gamma_{j}(\omega_{i}|_{i=1}^{N},\theta^{[0]})  \nonumber\\
& & \hspace*{2.0cm}  > 0
\label{eqn:a23}
\end{eqnarray}

To simplify expressions, in the remaining of this proof, the dependence on the parameter vector $\theta^{[0]}$ is omitted for all the matrices $\Gamma(\omega_{i}|_{i=1}^{N},\theta^{[0]})$, $\Gamma_{r,1}(\omega_{i}|_{i=1}^{N},\theta^{[0]})$, $\Gamma_{r,2}(\omega_{i}|_{i=1}^{N},\theta^{[0]})$, $\Gamma_{j}(\omega_{i}|_{i=1}^{N},\theta^{[0]})$, $\overline{\Pi}_{r}(\omega_{i}, \theta^{[0]})$ and $\overline{\Pi}_{j}(\omega_{i}, \theta^{[0]})$.

Define matrices $\widehat{\Gamma}_{r,1}^{[11]}(\omega_{i}|_{i=1}^{N})$, $\widehat{\Gamma}_{r,1}^{[21]}(\omega_{i}|_{i=1}^{N})$ and $\widehat{\Gamma}_{r,1}^{[22]}(\omega_{i}|_{i=1}^{N})$ respectively as
\begin{eqnarray*}
& & \hspace*{-1.0cm}
\widehat{\Gamma}_{r,1}^{[11]}(\omega_{i}|_{i=1}^{N}) =
(N-1)\left\{\left(I\otimes\overline{\Pi}^{T}_{r}(\omega_{1})\right)U_{\Psi,1}
U^{T}_{\Psi,1}\left(I\otimes\overline{\Pi}_{r}(\omega_{1})\right) \right\} \\
& & \hspace*{-1.0cm}
\widehat{\Gamma}_{r,1}^{[21]}(\omega_{i}|_{i=1}^{N}) =
-\left[\begin{array}{c}
\left(I\otimes\overline{\Pi}^{T}_{r}(\omega_{2})\right)U_{\Psi,1}
U^{T}_{\Psi,1}\left(I\otimes\overline{\Pi}_{r}(\omega_{1})\right)|_{i=2}^{N} \\
\left(I\otimes\overline{\Pi}^{T}_{r}(\omega_{3})\right)U_{\Psi,1}
U^{T}_{\Psi,1}\left(I\otimes\overline{\Pi}_{r}(\omega_{1})\right)|_{i=2}^{N} \\
\vdots \\
\left(I\otimes\overline{\Pi}^{T}_{r}(\omega_{N})\right)U_{\Psi,1}
U^{T}_{\Psi,1}\left(I\otimes\overline{\Pi}_{r}(\omega_{1})\right)|_{i=2}^{N}
\end{array}\right] \\
& & \hspace*{-1.0cm}
\widehat{\Gamma}_{r,1}^{[22]}(\omega_{i}|_{i=1}^{N}) =
{\rm\bf diag}\left\{\left(I\otimes\overline{\Pi}^{T}_{r}(\omega_{i})\right)U_{\Psi,1}
U^{T}_{\Psi,1}\left(I\otimes\overline{\Pi}_{r}(\omega_{i})\right)|_{i=2}^{N} \right\}
\end{eqnarray*}
Then from the definitions of the associated matrices, it is straightforward to show that
{\small
\begin{eqnarray*}
& & \hspace*{-0.9cm}
\Gamma^{T}_{r,1}(\omega_{i}|_{i=1}^{N}) \Gamma_{r,1}(\omega_{i}|_{i=1}^{N})
\!\!=\!\!\left[\!\!\begin{array}{cc}
\widehat{\Gamma}_{r,1}^{[11]}(\omega_{i}|_{i=1}^{N}) & \widehat{\Gamma}_{r,1}^{[21]T}(\omega_{i}|_{i=1}^{N}) \\
\widehat{\Gamma}_{r,1}^{[21]}(\omega_{i}|_{i=1}^{N}) &
\widehat{\Gamma}_{r,1}^{[22]}(\omega_{i}|_{i=1}^{N})
\end{array}\!\!\right]
\\
& & \hspace*{-0.9cm}
\Gamma^{T}_{r,2}(\omega_{i}|_{i=1}^{N})  \Gamma_{r,2}(\omega_{i}|_{i=1}^{N}) =
{\rm\bf diag}\left\{\left(I\otimes\overline{\Pi}^{T}_{r}(\omega_{i})\right)U_{\Psi,2}
U^{T}_{\Psi,2}\times \right. \\
& & \hspace*{5.0cm} \left.\left(I\otimes\overline{\Pi}_{r}(\omega_{i})\right)|_{i=1}^{N} \right\}
\\
& & \hspace*{-0.9cm}
\Gamma^{T}_{j}(\omega_{i}|_{i=1}^{N})
\Gamma_{j}(\omega_{i}|_{i=1}^{N}) =
{\rm\bf diag}\left\{\left(I\otimes\overline{\Pi}^{T}_{j}(\omega_{i})\right)
\left(I\otimes\overline{\Pi}_{j}(\omega_{i})\right)|_{i=1}^{N} \right\}
\end{eqnarray*} }

Substitute these equalities into Eq.(\ref{eqn:a23}), we have that
\begin{equation}
\hspace*{-1.4cm} \Gamma^{T}(\omega_{i}|_{i=1}^{N}) \Gamma(\omega_{i}|_{i=1}^{N}) \!=\!\left[\!\!\begin{array}{cc}
\overline{\Gamma}_{11}(\omega_{i}|_{i=1}^{N}) &
\widehat{\Gamma}_{r,1}^{[21]T}(\omega_{i}|_{i=1}^{N}) \\
\widehat{\Gamma}_{r,1}^{[21]}(\omega_{i}|_{i=1}^{N}) &
\overline{\Gamma}_{22}(\omega_{i}|_{i=1}^{N})
\end{array}\!\!\right]
\label{eqn:a24}
\end{equation}
in which
{\small
\begin{eqnarray*}
& & \hspace*{-0.6cm}
\overline{\Gamma}_{11}(\omega_{i}|_{i=1}^{N}) \!=\!
(N-1)\left(I\otimes\overline{\Pi}^{T}_{r}(\omega_{1})\right)U_{\Psi,1}
U^{T}_{\Psi,1}\left(I\otimes\overline{\Pi}_{r}(\omega_{1})\right) + \\
& & \hspace*{-0.3cm}\left(I\otimes\overline{\Pi}^{T}_{r}(\omega_{1})\right)U_{\Psi,2}
U^{T}_{\Psi,2}\left(I\otimes\overline{\Pi}_{r}(\omega_{1})\right) + \left(I\otimes\overline{\Pi}^{T}_{j}(\omega_{1})\right)\left(I\otimes\overline{\Pi}_{j}(\omega_{1})\right)
\\
& & \hspace*{-0.6cm}
\overline{\Gamma}_{22}(\omega_{i}|_{i=1}^{N}) \!=\!
{\rm\bf diag}\left\{\!\!\left(\!I\otimes\overline{\Pi}^{T}_{r}(\omega_{i})\!\right)(U_{\Psi,1}
U^{T}_{\Psi,1} \!+\! U_{\Psi,2}
U^{T}_{\Psi,2})\left(I\otimes\overline{\Pi}_{r}(\omega_{i})\right)\!+ \right.  \\
& & \hspace*{4.5cm} 
\left.\left(I\otimes\overline{\Pi}^{T}_{j}(\omega_{i})\right)
\left(I\otimes\overline{\Pi}_{j}(\omega_{i})\right)|_{i=2}^{N} \right\}
\end{eqnarray*} }

From the definitions of the matrices $U_{\Psi,1}$ and $U_{\Psi,2}$, which are given by Eq.(\ref{eqn:par-vec-2}), we have that the matrix $[U_{\Psi,1}\;\;  U_{\Psi,2}]$ is orthogonal, which means that $U_{\Psi,1}U^{T}_{\Psi,1}+U_{\Psi,2}
U^{T}_{\Psi,2} = I$ and $U^{T}_{\Psi,1}U_{\Psi,1}= I$, in which the identity matrices are of different dimensions. On the other hand, recall that the TFM $I - P(\theta) G_{zv}(\lambda)$ is guaranteed to be invertible when $\theta = \theta^{[0]}$, by the regularity and well-posedness assumptions A1) and A2). In addition, the MVP $V_{yv}^{[inv,2]}(\lambda)$ is of FRR for each $\lambda \in {\cal C}$. Hence the definition of the MVF $\Pi(\lambda,\theta)$ given by Eq.(\ref{eqn:pi}) implies that $\Pi(j\omega,\theta^{[0]})$ is of FCR at each $\omega \in {\cal R}$. This further means that at each $i=1,2,\cdots,N$, the matrix ${\rm\bf col}\{\overline{\Pi}_{r}(\omega_{i}),\;\; \overline{\Pi}_{j}(\omega_{i}) \}$ is of FCR. Hence the matrix $\overline{\Pi}^{T}_{r}(\omega_{i}) \overline{\Pi}_{r}(\omega_{i}) +  \overline{\Pi}^{T}_{j}(\omega_{i})\overline{\Pi}_{j}(\omega_{i})$ is always invertible.

On the basis of these observations, we further have that
\begin{eqnarray}
& & \hspace*{-1.2cm}
\overline{\Gamma}_{22}(\omega_{i}|_{i=1}^{N}) \!=\!
{\rm\bf diag}\left\{\left(I\otimes\overline{\Pi}^{T}_{r}(\omega_{i})\right)\left(I\otimes\overline{\Pi}_{r}(\omega_{i})\right) \right. + \nonumber\\
& & \hspace*{3.0cm}
\left.\left.\left(I\otimes\overline{\Pi}^{T}_{j}(\omega_{i})\right)
\left(I\otimes\overline{\Pi}_{j}(\omega_{i})\right)\right|_{i=2}^{N} \right\} \nonumber\\
& & \hspace*{0.3cm}
=\!{\rm\bf diag}\left\{\left.\left[I\otimes\left(\overline{\Pi}^{T}_{r}(\omega_{i})\overline{\Pi}_{r}(\omega_{i})+
\overline{\Pi}^{T}_{j}(\omega_{i})\overline{\Pi}_{j}(\omega_{i})\right)\right]\right|_{i=2}^{N} \right\} \nonumber\\
& & \hspace*{0.3cm}
>0
\end{eqnarray}
which implies that the matrix $\overline{\Gamma}_{22}(\omega_{i}|_{i=1}^{N})$ is always invertible. In addition, using the conclusions of Lemma \ref{lemma:4}, the following relation is obtained,
{\scriptsize
\begin{eqnarray}
& & \hspace*{-0.4cm}
\overline{\Gamma}_{11}(\omega_{i}|_{i=1}^{N}) -
\widehat{\Gamma}_{r,1}^{[21]T}(\omega_{i}|_{i=1}^{N}) \overline{\Gamma}^{-1}_{22}(\omega_{i}|_{i=1}^{N})
\widehat{\Gamma}_{r,1}^{[21]}(\omega_{i}|_{i=1}^{N}) \nonumber \\
& & \hspace*{-0.6cm}
\!=\!
(N \!-\! 1)\left(\!I\!\otimes\!\overline{\Pi}^{T}_{r}(\omega_{1})\!\right)U_{\Psi,1}
U^{T}_{\Psi,1}\left(\!I\!\otimes\!\overline{\Pi}_{r}(\omega_{1})\!\right) \!+\! \left(\!I\!\otimes\!\overline{\Pi}^{T}_{r}(\omega_{1})\!\right)U_{\Psi,2}U^{T}_{\Psi,2} \!\times \nonumber\\
& & \hspace*{-0.1cm}
\left(I\otimes\overline{\Pi}_{r}(\omega_{1})\right) +
\left(I\otimes\overline{\Pi}^{T}_{j}(\omega_{1})\right)\left(I\otimes\overline{\Pi}_{j}(\omega_{1})\right) -\ \sum_{i=2}^{N}\left\{\left(I\otimes\overline{\Pi}^{T}_{r}(\omega_{1})\right)\!\times \right.\nonumber \\
& & \hspace*{0.1cm}
\left.
U_{\Psi,1}U^{T}_{\Psi,1}\left(I\otimes\overline{\Pi}_{r}(\omega_{i})\right)
\left[I\otimes\left(\overline{\Pi}^{T}_{r}(\omega_{i})\overline{\Pi}_{r}(\omega_{i})
\!+\! \overline{\Pi}^{T}_{j}(\omega_{i})\overline{\Pi}_{j}(\omega_{i})\right)\right]^{-1} \!\times \right.
\nonumber\\
& & \hspace*{3.3cm}
\left. \left(I\otimes\overline{\Pi}^{T}_{r}(\omega_{i})\right)U_{\Psi,1}U^{T}_{\Psi,1}
\left(I\otimes\overline{\Pi}_{r}(\omega_{1})\right) \right\}  \nonumber \\
& & \hspace*{-0.6cm}
\!=\!
\left(I\otimes\overline{\Pi}^{T}_{j}(\omega_{1})\right)\left(I\otimes\overline{\Pi}_{j}(\omega_{1})\right)
+ \left(I\otimes\overline{\Pi}^{T}_{r}(\omega_{1})\right)U_{\Psi,2}U^{T}_{\Psi,2}\left(I\otimes\overline{\Pi}_{r}(\omega_{1})\right) \!+\!  \nonumber\\
& & \hspace*{-0.4cm}
(N\!-\!1)\left(\!I\!\otimes\!\overline{\Pi}^{T}_{r}(\omega_{1})\!\right)U_{\Psi,1}
U^{T}_{\Psi,1}\left(\!I\!\otimes\!\overline{\Pi}_{r}(\omega_{1})\!\right)
\!-\! \left(\!I\!\otimes\!\overline{\Pi}^{T}_{r}(\omega_{1})\right)U_{\Psi,1}U^{T}_{\Psi,1}\!\times \nonumber \\
& & \hspace*{-0.0cm}
\sum_{i=2}^{N}\left\{
I\otimes\left[\overline{\Pi}_{r}(\omega_{i})\left(\overline{\Pi}^{T}_{r}(\omega_{i})\overline{\Pi}_{r}(\omega_{i})
+\overline{\Pi}^{T}_{j}(\omega_{i})\overline{\Pi}_{j}(\omega_{i})\right)^{-1}
\overline{\Pi}^{T}_{r}(\omega_{i})\right]\right\}\times \nonumber \\
& & \hspace*{5.0cm}
U_{\Psi,1}U^{T}_{\Psi,1}
\left(I\otimes\overline{\Pi}_{r}(\omega_{1})\right) \nonumber \\
& & \hspace*{-0.6cm}
\!=\!
\left(I\otimes\overline{\Pi}^{T}_{j}(\omega_{1})\right)\left(I\otimes\overline{\Pi}_{j}(\omega_{1})\right)
+ \left(I\otimes\overline{\Pi}^{T}_{r}(\omega_{1})\right)U_{\Psi,2}U^{T}_{\Psi,2}\left(I\otimes\overline{\Pi}_{r}(\omega_{1})\right) \!+  \nonumber\\
& & \hspace*{0.4cm}
\left(I\otimes\overline{\Pi}^{T}_{r}(\omega_{1})\right)U_{\Psi,1}
U^{T}_{\Psi,1}
\sum_{i=2}^{N}\left\{I-
I\otimes\left[\overline{\Pi}_{r}(\omega_{i})\left(\overline{\Pi}^{T}_{r}(\omega_{i})\overline{\Pi}_{r}(\omega_{i})
\; + \right.\right.\right.\nonumber \\
& & \hspace*{1.4cm}
\left.\left.\left.
\overline{\Pi}^{T}_{j}(\omega_{i})\overline{\Pi}_{j}(\omega_{i})\right)^{-1}
\overline{\Pi}^{T}_{r}(\omega_{i})\right]\right\}
U_{\Psi,1}U^{T}_{\Psi,1}
\left(I\otimes\overline{\Pi}_{r}(\omega_{1})\right) \nonumber \\
& & \hspace*{-0.6cm}
\!=\!
\left(\!I\!\otimes\!\overline{\Pi}_{j}(\omega_{1})\!\right)^{\!T}\!\!\left(\!I\!\otimes\!\overline{\Pi}_{j}(\omega_{1})\!\right)
\!+\! \left[U^{T}_{\Psi,2}\left(\!I\! \otimes\! \overline{\Pi}_{r}(\omega_{1})\!\right)\!\right]^{\!T}\!\left[U^{T}_{\Psi,2}\left(\!I\otimes\!\overline{\Pi}_{r}(\omega_{1})\!\right) \!\right]\!+  \nonumber\\
& & \hspace*{0.2cm}
\sum_{i=2}^{N}\left\{\left[\left(I\otimes\left[U_{\Pi,2r}(\omega_{i})\;\;U_{\Pi,2j}(\omega_{i})\right]^{T}\right)
U_{\Psi,1}U^{T}_{\Psi,1}
\left(I\otimes\overline{\Pi}_{r}(\omega_{1})\right)\right]^{T}\times \right. \nonumber \\
& & \hspace*{0.4cm}
\left.\left[\left(I\otimes\left[U_{\Pi,2r}(\omega_{i})\;\;U_{\Pi,2j}(\omega_{i})\right]^{T}\right)
U_{\Psi,1}U^{T}_{\Psi,1}
\left(I\otimes\overline{\Pi}_{r}(\omega_{1})\right)\right]\right\}
\label{eqn:a25}
\end{eqnarray} }

Note that the requirement that the matrix $\Gamma(\omega_{i}|_{i=1}^{N})$ is of FCR is equivalent to the requirement that the matrix $\Gamma^{T}(\omega_{i}|_{i=1}^{N})\Gamma(\omega_{i}|_{i=1}^{N})$ is positive definite. If the matrix $\overline{\Pi}_{r}(\omega_{1})$ is of FCR. It is obvious from the Schur complement theorem \cite{hj1991}, Eq.(\ref{eqn:a24}) and the last equality of the above equation that, the matrix $\Gamma^{T}(\omega_{i}|_{i=1}^{N})\Gamma(\omega_{i}|_{i=1}^{N})$ is positive definite. Hence, the matrix $\Gamma(\omega_{i}|_{i=1}^{N})$ is of FCR. On the other hand, when the matrix $\overline{\Pi}_{j}(\omega_{1})$ is not of FCR, it can also be declared from Eqs.(\ref{eqn:a24}) and (\ref{eqn:a25}), as well as the aforementioned equivalence, that the matrix $\Gamma(\omega_{i}|_{i=1}^{N})$ is of FCR, if and only if the following matrix is of FCR,
\begin{displaymath}
\left[\begin{array}{c}
I\otimes\overline{\Pi}_{j}(\omega_{1}) \\
U^{T}_{\Psi,2}\left(I\otimes\overline{\Pi}_{r}(\omega_{1})\right) \\
\left(I\otimes\left[\begin{array}{c} U^{T}_{\Pi,2r}(\omega_{2})\\ U^{T}_{\Pi,2j}(\omega_{2})\end{array} \right]\right)
U_{\Psi,1}U^{T}_{\Psi,1}
\left(I\otimes\overline{\Pi}_{r}(\omega_{1})\right) \\
\vdots \\
\left(I\otimes\left[\begin{array}{c} U^{T}_{\Pi,2r}(\omega_{N}) \\ U^{T}_{\Pi,2j}(\omega_{N})\end{array}\right]\right)
U_{\Psi,1}U^{T}_{\Psi,1}
\left(I\otimes\overline{\Pi}_{r}(\omega_{1})\right)
\end{array}\right]
\end{displaymath}

According to Lemma \ref{lemma:0} and properties of Kronecker matrix products, the latter is equivalent to that the next matrix is of FCR,
\begin{displaymath}
\left[\begin{array}{c}
U^{T}_{\Psi,2}\left(I\otimes\overline{\Pi}_{r}(\omega_{1})\right) \\
\left(I\otimes\left[\begin{array}{c} U^{T}_{\Pi,2r}(\omega_{2})\\ U^{T}_{\Pi,2j}(\omega_{2})\end{array} \right]\right)
U_{\Psi,1}U^{T}_{\Psi,1}
\left(I\otimes\overline{\Pi}_{r}(\omega_{1})\right) \\
\vdots \\
\left(I\otimes\left[\begin{array}{c} U^{T}_{\Pi,2r}(\omega_{N}) \\ U^{T}_{\Pi,2j}(\omega_{N})\end{array}\right]\right)
U_{\Psi,1}U^{T}_{\Psi,1}
\left(I\otimes\overline{\Pi}_{r}(\omega_{1})\right)
\end{array}\right]\left(I\otimes\overline{\Pi}_{j,r}^{\perp}(\omega_{1})\right)
\end{displaymath}
which is further is equivalent to that the following matrix is of FCR,
\begin{equation}
\left[\begin{array}{c}
U^{T}_{\Psi,2} \\
\left(I\otimes\left[\begin{array}{c} U^{T}_{\Pi,2r}(\omega_{2})\\ U^{T}_{\Pi,2j}(\omega_{2})\end{array} \right]\right)
U_{\Psi,1}U^{T}_{\Psi,1} \\
\vdots \\
\left(I\otimes\left[\begin{array}{c} U^{T}_{\Pi,2r}(\omega_{N}) \\ U^{T}_{\Pi,2j}(\omega_{N})\end{array}\right]\right)
U_{\Psi,1}U^{T}_{\Psi,1}
\end{array}\right] \left(I\otimes\left[\overline{\Pi}_{r}(\omega_{1})
\overline{\Pi}_{j,r}^{\perp}(\omega_{1})\right]\right)
\label{eqn:a27}
\end{equation}

To guarantee that the above matrix is of FCR, it is obvious that the matrix $I\otimes\left[\overline{\Pi}_{r}(\omega_{1})
\overline{\Pi}_{j,r}^{\perp}(\omega_{1})\right]$ must be FCR. According to the properties of matrix Kronecker products, the latter is equivalent to that the matrix $\overline{\Pi}_{r}(\omega_{1})
\overline{\Pi}_{j,r}^{\perp}(\omega_{1})$ is of FCR.

From the definition of the matrix $\Xi(\omega_{1})$, it is obvious that the matrix $\left\{I\otimes\Xi(\omega_{1})\right\}$ is of FRR. In addition,
\begin{displaymath}
\left\{I \otimes \Xi(\omega_{1})\right\}
\left\{I \otimes \left[\overline{\Pi}_{r}(\omega_{1})
\overline{\Pi}_{j,r}^{\perp}(\omega_{1})\right]\right\}
=
I\otimes\left[\Xi(\omega_{1}) \overline{\Pi}_{r}(\omega_{1})
\overline{\Pi}_{j,r}^{\perp}(\omega_{1})\right]
=
0
\end{displaymath}

On the basis of these properties of the matrix $\Xi(\omega_{1})$ and Lemma \ref{lemma:0}, we have that the matrix of Eq.(\ref{eqn:a27}) is of FCR, if and only if the following matrix is of FCR,
\begin{displaymath}
\left[\begin{array}{c}
U^{T}_{\Psi,2} \\
I\otimes \Xi(\omega_{1})\\
\left(I\otimes\left[\begin{array}{c} U^{T}_{\Pi,2r}(\omega_{2})\\ U^{T}_{\Pi,2j}(\omega_{2})\end{array} \right]\right)
U_{\Psi,1}U^{T}_{\Psi,1} \\
\vdots \\
\left(I\otimes\left[\begin{array}{c} U^{T}_{\Pi,2r}(\omega_{N}) \\ U^{T}_{\Pi,2j}(\omega_{N})\end{array}\right]\right)
U_{\Psi,1}U^{T}_{\Psi,1}
\end{array}\right]
\end{displaymath}

Recall that $U^{T}_{\Psi,2}U_{\Psi,1} =0$ and $U^{T}_{\Psi,1}U_{\Psi,1} = I$. It can be straightforwardly shown, once again on the basis of Lemma \ref{lemma:0}, that the above matrix is of FCR, if and only if
the following matrix is of FCR,
\begin{displaymath}
\left[\begin{array}{c}
I\otimes \Xi(\omega_{1})\\
\left(I\otimes\left[\begin{array}{c} U^{T}_{\Pi,2r}(\omega_{2})\\ U^{T}_{\Pi,2j}(\omega_{2})\end{array} \right]\right)
U_{\Psi,1}U^{T}_{\Psi,1} \\
\vdots \\
\left(I\otimes\left[\begin{array}{c} U^{T}_{\Pi,2r}(\omega_{N}) \\ U^{T}_{\Pi,2j}(\omega_{N})\end{array}\right]\right)
U_{\Psi,1}U^{T}_{\Psi,1}
\end{array}\right]U_{\Psi,1}
\end{displaymath}
which is further equivalent to that the following matrix is of FCR,
\begin{displaymath}
\left[\begin{array}{c}
\left(I\otimes \Xi(\omega_{1})\right) U_{\Psi,1}\\
\left(I\otimes\left[\begin{array}{c} U^{T}_{\Pi,2r}(\omega_{2})\\ U^{T}_{\Pi,2j}(\omega_{2})\end{array} \right]\right)
U_{\Psi,1} \\
\vdots \\
\left(I\otimes\left[\begin{array}{c} U^{T}_{\Pi,2r}(\omega_{N}) \\ U^{T}_{\Pi,2j}(\omega_{N})\end{array}\right]\right)
U_{\Psi,1}
\end{array}\right]
\end{displaymath}

Through an direct application of Lemma \ref{lemma:0}, it can be proven that the latter is equivalent to that the following matrix is of FCR, noting that $U^{T}_{\Psi,2}U_{\Psi,1} =0$ and $U_{\Psi,1}$ is of FCR.
\begin{displaymath}
\left[\begin{array}{c}
U^{T}_{\Psi,2} \\
I \otimes {\left[\begin{array}{c}
\Xi(\omega_{1}) \\
{[U_{\Pi,2r}(\omega_{2}) \;\; U_{\Pi,2j}(\omega_{2})]^{T}} \\
\vdots \\
{[U_{\Pi,2r}(\omega_{N}) \;\; U_{\Pi,2j}(\omega_{N})]^{T}}
\end{array} \right]}
\end{array}\right]
\end{displaymath}

This completes the proof.   \hspace{\fill}$\Diamond$

\noindent\textbf{Proof of Lemma \ref{lemma:5}:} From the definition of the right coprime factorization of a TFM \cite{Kailath1980,zdg1996}, it can be directly declared that $D(\lambda)$ is square and invertible at each $\lambda = j\omega$ with $\omega\in {R}$ for a continuous time system, and $\lambda = e^{j\omega}$ with $\omega\in {R}$ for a discrete time system.  Denote the real and imaginary parts of $D^{-1}(j\omega)$ respectively by $D^{[inv]}_{r}(\omega)$ and $D^{[inv]}_{j}(\omega)$. Then we have that the following matrix is invertible
\begin{equation}
\left[\begin{array}{rr}
D^{[inv]}_{r}(\omega) & -D^{[inv]}_{j}(\omega) \\
D^{[inv]}_{j}(\omega) & D^{[inv]}_{r}(\omega) \end{array}\right]
\label{eqn:a26}
\end{equation}
and
\begin{eqnarray}
\Pi(j\omega) &=& \Pi_{r}(\omega) + j \Pi_{j}(\omega) \nonumber \\
&=& \left[N_{r}(\omega)+ j N_{j}(\omega)\right]\times \left[D^{[inv]}_{r}(\omega)+ j D^{[inv]}_{j}(\omega)\right] \nonumber\\
&=& \left[N_{r}(\omega)D^{[inv]}_{r}(\omega) - N_{j}(\omega)D^{[inv]}_{j}(\omega)\right] + \nonumber \\
& & \hspace*{1cm} j \left[N_{j}(\omega)D^{[inv]}_{r}(\omega)+ N_{r}(\omega) D^{[inv]}_{j}(\omega)\right]
\end{eqnarray}
Hence
\begin{equation}
\left[\!\!\begin{array}{rr} \Pi_{r}(\omega) & -\Pi_{j}(\omega)\\
 \Pi_{j}(\omega) & \Pi_{r}(\omega) \end{array}\!\!\right]
\!=\!
\left[\!\!\begin{array}{rr} N_{r}(\omega) & -N_{j}(\omega) \\
 N_{j}(\omega) & N_{r}(\omega) \end{array}\!\!\right]
\left[\!\!\begin{array}{rr} D^{[inv]}_{r}(\omega) & -D^{[inv]}_{j}(\omega) \\
 D^{[inv]}_{j}(\omega) & D^{[inv]}_{r}(\omega) \end{array}\!\!\right]
\end{equation}

From this relation and the invertibility of the matrix in Eq.(\ref{eqn:a26}), it can be straightforwardly shown that
\begin{equation}
\left[\!\!\begin{array}{rr}
 \Pi_{j}(\omega) & \Pi_{r}(\omega) \end{array}\!\!\right]_{r}^{\perp}
\!=\!
\left[\!\!\begin{array}{rr} D^{[inv]}_{r}(\omega) & -D^{[inv]}_{j}(\omega) \\
 D^{[inv]}_{j}(\omega) & D^{[inv]}_{r}(\omega) \end{array}\!\!\right]^{-1}
 \left[\!\!\begin{array}{rr}  N_{j}(\omega) & N_{r}(\omega) \end{array}\!\!\right]_{r}^{\perp}
\end{equation}
which further leads to that
\begin{eqnarray}
& & \left[ \Pi_{r}(\omega)\;\; -\Pi_{j}(\omega) \right] \left[ \Pi_{j}(\omega)\;\; \Pi_{r}(\omega) \right]_{r}^{\perp} \nonumber\\
&=&
\left(\left[\!\!\begin{array}{rr} N_{r}(\omega) & -N_{j}(\omega) \end{array}\!\!\right]
\left[\!\!\begin{array}{rr} D^{[inv]}_{r}(\omega) & -D^{[inv]}_{j}(\omega) \\
 D^{[inv]}_{j}(\omega) & D^{[inv]}_{r}(\omega) \end{array}\!\!\right]\right) \times \nonumber\\
& &
\left(\left[\!\!\begin{array}{rr} D^{[inv]}_{r}(\omega) & -D^{[inv]}_{j}(\omega) \\
 D^{[inv]}_{j}(\omega) & D^{[inv]}_{r}(\omega) \end{array}\!\!\right]^{-1}
 \left[\!\!\begin{array}{rr}  N_{j}(\omega) & N_{r}(\omega) \end{array}\!\!\right]_{r}^{\perp}\right) \nonumber\\
&=&
\left[ N_{r}(\omega)\;\; -N_{j}(\omega) \right] \left[ N_{j}(\omega)\;\; N_{r}(\omega) \right]_{r}^{\perp}
\end{eqnarray}
and therefore completes the proof.   \hspace{\fill}$\Diamond$

\noindent\textbf{Proof of Theorem \ref{theo:3}:} From the definitions of the MVPs $\overline{N}_{r}(\lambda)$ and $\overline{N}_{j}(\lambda)$, it is obvious that at each $\omega \in {\cal R}$, the following equality is valid
\begin{equation}
N(j\omega) = \overline{N}_{r}(\omega) + j \overline{N}_{j}(\omega)
\end{equation}

On the other hand, from the constructions of the MVP $V^{[inv,2]}_{\overline{N}}(\lambda)$, we have that it is of FRR at each $\lambda \in {\cal C}$ and
\begin{eqnarray}
\left[\overline{N}_{j}(\lambda)\;\; \overline{N}_{r}(\lambda)\right]V^{[inv,2]T}_{\overline{N}}(\lambda)
&=& U_{\overline{N},1}(\lambda) \Sigma_{\overline{N}}(\lambda)V^{T}_{\overline{N},1}(\lambda)V^{[inv,2]T}_{\overline{N}}(\lambda) \nonumber\\
&=& U_{\overline{N},1}(\lambda) \Sigma_{\overline{N}}(\lambda)\left(V^{[inv,2]}_{\overline{N}}(\lambda)V_{\overline{N},1}(\lambda)\right)^{T} \nonumber\\
&=& 0
\end{eqnarray}
These mean that at each $\omega \in {\cal R}$, the column vectors of the matrix  $V^{[inv,2]T}_{\overline{N}}(\omega)$ consist of a basis of the right null space of the matrix $\left[\overline{N}_{j}(\omega)\;\; \overline{N}_{r}(\omega)\right]$.

Using similar arguments, it can be shown that for every $\omega \in {\cal R}$, the row vectors of the matrix   $U^{[inv,2]}_{\widetilde{N}}(\omega)$  consist of a basis of the left null space of the matrix $\left[\overline{N}_{r}(\omega)\;\; -\overline{N}_{j}(\omega)\right]\left[\overline{N}_{j}(\omega)\;\; \overline{N}_{r}(\omega)\right]_{r}^{\perp}$.

The proof can now be completed using the conclusions of Lemma \ref{lemma:5}.   \hspace{\fill}$\Diamond$

\noindent\textbf{Proof of Theorem \ref{theo:4}:} From Eqs. (\ref{eqn:pi}) and (\ref{eqn:in-out}), we have that
\begin{eqnarray}
& & U^{T}_{out}(-\lambda,\theta)U^{T}_{in}(-\lambda,\theta)\Pi(\lambda,\theta) \nonumber\\
&=& \left[U_{in}(-\lambda,\theta)U_{out}(-\lambda,\theta)\right]^{T}\Pi(\lambda,\theta) \nonumber\\
&=& \left\{
\left[ I - P(\theta)G_{zv}(\lambda) \right]^{-T}V_{yv,1}(\lambda)\right\}^{T}\left\{\left[ I - P(\theta)G_{zv}(\lambda) \right]V^{[inv,2]T}_{yv}(\lambda)\right\} \nonumber\\
&=& \left\{V^{[inv,2]}_{yv}(\lambda)V_{yv,1}(\lambda)\right\}^{T} \nonumber\\
&=& 0
\end{eqnarray}

Recall that at each $\lambda = j\omega$, an outer function is square and invertible \cite{zdg1996}. Note also that both the TFM $U_{in}(\lambda,\theta)$ and the TFM $U_{out}(-\lambda,\theta)$ are real MVFs. We therefore have that at each $\omega \in {\cal R}$,
\begin{equation}
U^{H}_{in}(j\omega,\theta)\Pi(j\omega,\theta) = 0
\end{equation}

On the other hand, recall that at each $\lambda = j\omega$, the product of the conjugate transpose of an inner MVF and itself is equal to an identity matrix \cite{zdg1996}. This implies that $U_{in}(j\omega,\theta)$ is of FCR and
$U^{H}_{in}(j\omega,\theta)U_{in}(j\omega,\theta) = I$.

The proof can now be completed through comparing dimensions of the matrix $U_{in}(j\omega,\theta)$ and the matrix $U_{\Pi,2}(j\omega,\theta)$.   \hspace{\fill}$\Diamond$

\noindent\textbf{Proof of Theorem \ref{theo:5}:} For an  arbitrary parameter vector $\theta \in {\rm\bf\Theta}$, define a TFM $\Delta_{H}(\lambda,\theta)$ and a real matrix $\Delta_{P}(\theta)$ respectively as
\begin{equation}
\Delta_{H}(\lambda,\theta) = H(\lambda,\theta) - H(\lambda,\theta^{[0]}), \hspace{0.25cm}
\Delta_{P}(\theta) = \sum_{k=1}^{q} (\theta_{k}-\theta_{k}^{[0]}) P_{k}
\end{equation}
Then from the definition of the TFM ${H}(\lambda,\theta)$, it can be straightforwardly shown that
\begin{eqnarray}
& & \hspace*{-1.5cm} \Delta_{H}(\lambda,\theta) \!=\! G_{yv}(\lambda) \left[I \!-\! P(\theta^{[0]}) G_{zv}(\lambda) \right]^{-1}\!\!
\Delta_{P}(\theta) \left\{ I \!-\! \left[I \!-\!  G_{zv}(\lambda)  \times \right.\right.  \nonumber \\
& & \left.\left. \hspace*{-0.2cm} P(\theta^{[0]}) \right]^{-1}\!\! G_{zv}(\lambda)\Delta_{P}(\theta) \right\}^{-1}\!\! \left[I \!-\!  G_{zv}(\lambda) P(\theta^{[0]}) \right]^{-1}\!\! G_{zu}(\lambda)
\label{eqn:mod-err}
\end{eqnarray}

Assume that $\omega_{i} \in {\cal R}$ in which $i=1,2,\cdots ,N$, and $\omega_{i}$ is different from $\omega_{j}$ whenever $i \neq j$. Moreover, assume that $\Delta_{i}$ with $i=1, 2, \cdots, N$, is a complex matrix with a dimension compatible with that of the TFM $\Delta_{H}(\lambda,\theta)$. Then according to Lemma \ref{lemma:0-orth} and Eq.(\ref{eqn:mod-err}), there exists a parameter vector $\theta$ in the set ${\rm\bf\Theta}$, such that $\Delta_{H}(j\omega_{i},\theta) = \Delta_{i}$ for each $i=1,2,\cdots ,N$, only when the complex matrix $\Delta_{i}$ satisfies the following conditions simultaneously,
\begin{equation}
U_{yv}^{[inv,2]}(j\omega_{i}) \Delta_{i} = 0, \hspace{0.25cm}
\Delta_{i} V_{zu}^{[inv,2]T}(j\omega_{i}) = 0
\end{equation}
These further imply that there exists one and only one complex matrix $\overline{\Delta}_{i}$, such that
\begin{equation}
\Delta_{i} = U_{yv,1}(j\omega_{i}) \overline{\Delta}_{i} V_{zu,1}^{T}(j\omega_{i})
\label{eqn:data-err}
\end{equation}
recalling that at each $\omega \in {\cal R}$, both the matrix $U_{yv}^{[inv,2]}(j\omega) $ and the matrix $V_{zu}^{[inv,2]}(j\omega)$ are of FRR, and $U_{yv}^{[inv,2]}(j\omega) U_{yv,1}(j\omega) = 0  $ and $V_{zu}^{[inv,2]}(j\omega) V_{zu,1}(j\omega) = 0$.

Substitute this expression into Eq.(\ref{eqn:mod-err}) and note that both the matrix $U_{yv,1}(j\omega) $ and the matrix $V_{zu,1}(j\omega)$ are of FCR whenever $\omega \in {\cal R}$, while both the matrix $\Sigma_{yv}(j\omega)$ and the matrix $\Sigma_{zu}(j\omega)$ are square and invertible, we further have that if $\Delta_{H}(j\omega_{i},\theta) = \Delta_{i}$, then
\begin{eqnarray}
& & \hspace*{-1.5cm} \Sigma_{yv}^{-1}(j\omega_{i}) \overline{\Delta}_{i} \Sigma_{zu}^{-1}(j\omega_{i}) \!=\!  V_{yv,1}^{T}(j\omega_{i})\left[I \!-\! P(\theta^{[0]}) G_{zv}(j\omega_{i}) \right]^{-1}\!
\Delta_{P}(\theta)  \times   \nonumber \\
& & \hspace*{1cm} \left\{ I \!-\! \left[I \!-\!  G_{zv}(j\omega_{i}) P(\theta^{[0]}) \right]^{-1}\! G_{zv}(j\omega_{i})\Delta_{P}(\theta) \right\}^{-1} \times   \nonumber \\
& & \hspace*{1cm} \left[I -  G_{zv}(j\omega_{i}) P(\theta^{[0]}) \right]^{-1} U_{zu,1}(j\omega_{i})
\label{eqn:data-err-2}
\end{eqnarray}

According to Lemma \ref{lemma:1}, Eq.(\ref{eqn:data-err-2}) is equivalent to the following equality
\begin{eqnarray}
& & \hspace*{-1.0cm} V_{yv}^{[inv,1]T}(j\omega_{i})\Sigma_{yv}^{-1}(j\omega_{i}) \overline{\Delta}_{i} \Sigma_{zu}^{-1}(j\omega_{i}) U_{zu}^{[inv,1]}(j\omega_{i}) + A(j\omega_{i}) -  \nonumber \\
& & \hspace*{-1.0cm}
V_{yv}^{[inv,1]T}(j\omega_{i})V_{yv,1}^{T}(j\omega_{i})A(j\omega_{i})
U_{zu,1}(j\omega_{i})U_{zu}^{[inv,1]}(j\omega_{i}) \nonumber \\
& & \hspace*{-1.15cm}
\!=\!  \left[I \!-\! P(\theta^{[0]}) G_{zv}(j\omega_{i}) \right]^{-1}\!
\Delta_{P}(\theta) \! \left\{ I \!-\! \left[I \!-\!  G_{zv}(j\omega_{i}) P(\theta^{[0]}) \right]^{-1}\! \right. \times   \nonumber \\
& & \hspace*{-1.0cm} \left. G_{zv}(j\omega_{i})\Delta_{P}(\theta) \right\}^{-1}\!\! \left[I -  G_{zv}(j\omega_{i}) P(\theta^{[0]}) \right]^{-1}
\label{eqn:data-err-3}
\end{eqnarray}
in which $A(j\omega_{i})$ is an arbitrary complex matrix with a compatible dimension.

When the TFM $G_{zu}(\lambda)$ is of FNRR, the MVP $U_{zu,2}(\lambda)$ is empty and the MVP $U_{zu,1}(\lambda)$ is square and invertible at each $\lambda \in {\cal C}$. Under such a situation, the above equation can be further reduced to
\begin{eqnarray}
& & \hspace*{-1.0cm} V_{yv}^{[inv,1]T}(j\omega_{i})\Sigma_{yv}^{-1}(j\omega_{i}) \overline{\Delta}_{i} \Sigma_{zu}^{-1}(j\omega_{i}) U_{zu}^{[inv,1]}(j\omega_{i}) + \nonumber \\
& & \hspace*{-1.0cm}
\left[ I - V_{yv}^{[inv,1]T}(j\omega_{i})V_{yv,1}^{T}(j\omega_{i}) \right] A(j\omega_{i})
\nonumber \\
& & \hspace*{-1.15cm}
\!=\!  \left[I \!-\! P(\theta^{[0]}) G_{zv}(j\omega_{i}) \right]^{-1}\!
\Delta_{P}(\theta) \! \left\{ I \!-\! \left[I \!-\!  G_{zv}(j\omega_{i}) P(\theta^{[0]}) \right]^{-1}\! \right. \times   \nonumber \\
& & \hspace*{-1.0cm} \left. G_{zv}(j\omega_{i})\Delta_{P}(\theta) \right\}^{-1}\!\! \left[I -  G_{zv}(j\omega_{i}) P(\theta^{[0]}) \right]^{-1}
\label{eqn:data-err-4}
\end{eqnarray}

Denote $A(j\omega_{i})\left\{ I \!-\!   G_{zv}(j\omega_{i}) \left[ P(\theta^{[0]}) + \Delta_{P}(\theta) \right] \right\}$ by $\overline{A}(j\omega_{i})$. Note that $P(\theta^{[0]}) + \Delta_{P}(\theta) = P(\theta)$ according to their definitions. Moreover, the TFM $I \!-\!   G_{zv}(\lambda) P(\theta) $ is guaranteed to be invertible for each parameter vector $\theta \in {\rm\bf\Theta}$ from the regularity and well-posedness assumptions A1) and A2). It is obvious that the mapping between the complex matrices ${A}(j\omega_{i})$ and $\overline{A}(j\omega_{i})$ are bijective.

With these symbols, as well as the MVF $\Pi(\lambda, \theta)$ defined by Eq.(\ref{eqn:pi}), the relation between $\Delta_{P}(\theta)$ and $\overline{\Delta}_{i}$ given by Eq.(\ref{eqn:data-err-4}) can be equivalently rewritten as follows,
\begin{eqnarray}
& & \hspace*{-1.1cm} \Delta_{P}(\theta) \!=\! \Phi_{l}(j\omega_{i},\theta^{[0]}) \overline{\Delta}_{i}
\Phi_{r}(j\omega_{i},\theta^{[0]}) + \Pi(j\omega_{i},\theta^{[0]}) \overline{A}(j\omega_{i}) +  \nonumber \\
& & \hspace*{-0.00cm}
\Phi_{l}(j\omega_{i},\theta^{[0]}) \overline{\Delta}_{i} \Sigma_{zu}^{-1}(j\omega_{i}) U_{zu}^{[inv,1]}(j\omega_{i}) G_{zv} (j\omega_{i}) \Delta_{P}(\theta)
\label{eqn:data-err-5}
\end{eqnarray}
which can be further expressed equivalently as
\begin{eqnarray}
& & \hspace*{-0.90cm} \left[ I + \Phi_{l}(j\omega_{i},\theta^{[0]}) \overline{\Delta}_{i} \Sigma_{zu}^{-1}(j\omega_{i}) U_{zu}^{[inv,1]}(j\omega_{i}) G_{zv} (j\omega_{i}) \right] \Delta_{P}(\theta) \nonumber \\
& & \hspace*{-1.20cm}
= \Phi_{l}(j\omega_{i},\theta^{[0]}) \overline{\Delta}_{i}
\Phi_{r}(j\omega_{i},\theta^{[0]}) + \Pi(j\omega_{i},\theta^{[0]}) \overline{A}(j\omega_{i})
\label{eqn:data-err-6}
\end{eqnarray}

Note that the maximum singular value of a matrix and its Frobenius norm, that is, the square root of the sum of the square of its elements, are two equivalent norms \cite{hj1991,gv1989}. It can therefore be declared that when $\varepsilon$ is small enough, it is certain that for each $i=1,2,\cdots,N$, $\overline{\sigma}(\overline{\Delta}_{i}) << 1$. As the maximum singular values of the complex matrices $\Phi_{l}(j\omega,\theta)$,
$\Sigma_{zu}^{-1}(j\omega)$, $U_{zu}^{[inv,1]}(j\omega)$ and $G_{zv} (j\omega)$  are bounded for each $\omega \in {\cal R}$ and each $\theta \in {\rm\bf\Theta}$, it is obvious that
\begin{displaymath}
\overline{\sigma}\left( \Phi_{l}(j\omega_{i},\theta^{[0]}) \overline{\Delta}_{i} \Sigma_{zu}^{-1}(j\omega_{i}) U_{zu}^{[inv,1]}(j\omega_{i}) G_{zv} (j\omega_{i}) \right)  << 1
\end{displaymath}
Therefore, under such an assumption, Eq.(\ref{eqn:data-err-6}) can be  approximately written as
\begin{equation}
\hspace*{-0.60cm} \Delta_{P}(\theta) \approx \Phi_{l}(j\omega_{i},\theta^{[0]}) \overline{\Delta}_{i}
\Phi_{r}(j\omega_{i},\theta^{[0]}) + \Pi(j\omega_{i},\theta^{[0]}) \overline{A}(j\omega_{i})
\label{eqn:data-err-7}
\end{equation}
or equivalently as
\begin{eqnarray}
& & \hspace*{-0.90cm} \Delta_{P}(\theta) \!=\! \Phi_{l}(j\omega_{i},\theta^{[0]}) \overline{\Delta}_{i}
\Phi_{r}(j\omega_{i},\theta^{[0]}) + \Pi(j\omega_{i},\theta^{[0]}) \overline{A}(j\omega_{i}) + \nonumber \\
& & \hspace*{2.0cm} O\left( max \left\{\overline{\sigma}^{2}(\overline{\Delta}_{i}), \overline{\sigma}(\overline{\Delta}_{i})\times\overline{\sigma}(\overline{A}(j\omega_{i}))\right\}\right)
\label{eqn:data-err-7-a}
\end{eqnarray}

Denote the real and imaginary parts of the complex matrices  $\overline{A}(j\omega_{i})$ and $\overline{\Delta}_{i}$ respectively by $\overline{A}_{r}(\omega_{i})$, $\overline{A}_{j}(\omega_{i})$, $\overline{\Delta}_{i,r}$ and $\overline{\Delta}_{i,j}$. Moreover, denote ${\rm\bf vec}\left({\rm\bf col}\left\{\overline{A}_{r}(\omega_{i}),\; \overline{A}_{j}(\omega_{i})\right\}\right)$ and ${\rm\bf vec}\left({\rm\bf col}\left\{\overline{\Delta}_{i,r},\; \overline{\Delta}_{i,j}\right\}\right)$ by $\overline{a}(\omega_{i})$ and $\overline{\xi}_{Hi}$.
Vectorize both sides of Eq.(\ref{eqn:data-err-7-a}). Direct matrix operations show that Eq.(\ref{eqn:data-err-7-a}) is equal to the following two equations,
\small
\begin{eqnarray}
& & \hspace*{-1.40cm}
\Psi(\theta\!-\!\theta^{[0]}) \!=\! \left[ I \!\otimes\! \overline{\Pi}_{r}(\omega_{i},\theta^{[0]}) \right] \overline{a}(\omega_{i}) \!+\!  Q_{r}(\omega_{i},\theta^{[0]})\overline{\xi}_{Hi} \!+\! O\left( ||\overline{\xi}_{i}||_{2}^{2} \right)
\label{eqn:data-err-8} \\
& & \hspace*{-1.40cm}
\left[ I \otimes \overline{\Pi}_{j}(\omega_{i},\theta^{[0]})\right] \overline{a}(\omega_{i}) +  Q_{j}(\omega_{i},\theta^{[0]})\overline{\xi}_{Hi} = O\left( ||\overline{\xi}_{i}||_{2}^{2} \right)
\label{eqn:data-err-9}
\end{eqnarray}
Here, with a little abuse of terminology, $O\left( ||\overline{\xi}_{i}||_{2}^{2} \right)$ stands for
$O\left( max \left\{||\overline{\xi}_{Hi}||_{2}^{2}, ||\overline{\xi}_{Hi}||\times ||\overline{a}(\omega_{i})||\right\} \right)$, in order to simplify expressions.

On the basis of the SVD of the matrix $\Psi$, as well as Lemma \ref{lemma:1}, it can be declared that there is a parameter vector $\theta \in {\rm\bf\Theta}$, such that Eq.(\ref{eqn:data-err-8}) is satisfied, if and only if there exist a real vector $\overline{a}(\omega_{i})$ and a real vector $\overline{\xi}_{Hi}$, such that
\begin{equation}
\hspace*{-0.40cm}
U_{\Psi,2}^{T} \left\{ \left[ I \!\otimes\! \overline{\Pi}_{r}(\omega_{i},\theta^{[0]}) \right] \overline{a}(\omega_{i}) \!+\!  Q_{r}(\omega_{i},\theta^{[0]})\overline{\xi}_{Hi} \right\} \!=\! O\left( ||\overline{\xi}_{i}||_{2}^{2} \right)
\label{eqn:data-err-8-a}
\end{equation}
In addition, when this condition is satisfied and the matrix $\Psi$ is of FCR, the desirable parameter vector $\theta$ is uniquely determined by the associated vectors  $\overline{a}(\omega_{i})$ and $\overline{\xi}_{Hi}$ that can be expressed as
\begin{eqnarray}
& & \hspace*{-1.40cm}
\theta \!=\! \theta^{[0]} \!+\! V_{\Psi}\Sigma_{\Psi}^{-1}U_{\Psi,1}^{T} \left\{ \left[ I \!\otimes\! \overline{\Pi}_{r}(\omega_{i},\theta^{[0]}) \right] \overline{a}(\omega_{i}) \!+\!  Q_{r}(\omega_{i},\theta^{[0]})\overline{\xi}_{Hi} \right\} \!+\! \nonumber \\
& & \hspace*{5cm} O\left( ||\overline{\xi}_{i}||_{2}^{2} \right)
\label{eqn:data-err-8-b}
\end{eqnarray}

Denote the vectors ${\rm\bf col}\left\{\overline{a}(\omega_{i})|_{i=1}^{N} \right\}$ and ${\rm\bf col}\left\{\overline{\xi}_{Hi}|_{i=1}^{N} \right\}$ respectively by $\overline{a}$ and $\overline{\xi}_{H}$. Then on the basis of Eqs.(\ref{eqn:data-err-9}) and (\ref{eqn:data-err-8-b}), similar arguments as those in the proof of Theorem \ref{theo:1} show that, when $\sum_{i=1}^{N}||\Delta_{i}||_{F}^{2} \leq \varepsilon$ and $\varepsilon$ is small enough, there is a parameter $\theta$ in the set ${\rm\bf\Theta}$, such that $\Delta_{H}(j\omega_{i},\theta) = \Delta_{i}$ is satisfied at each $i=1,2,\cdots$, if and only if there exists a real vector $\overline{a}$, such that
\begin{equation}
\Gamma(\omega_{i}|_{i=1}^{N},\theta^{[0]}) \overline{a} +
\Omega(\omega_{i}|_{i=1}^{N},\theta^{[0]}) \overline{\xi}_{H}
=  O\left( ||\overline{\xi}||_{2}^{2} \right)
\label{eqn:data-err-10}
\end{equation}
Here, once again in order to simplify expressions, $O\left( ||\overline{\xi}||_{2}^{2} \right)$ is adopted to denote
$O\left( max \left\{||\overline{\xi}_{H}||_{2}^{2}, ||\overline{\xi}_{H}||\times ||\overline{a}||\right\} \right)$.

When the parameter vector value $\theta^{[0]}$ of the descriptor system $\rm\bf\Sigma$ is globally identifiable from its frequency responses at the frequencies $\omega_{i}|_{i=1}^{N}$, Theorem \ref{theo:1} reveals that the matrix $\Gamma(\omega_{i}|_{i=1}^{N},\theta^{[0]})$ is of FCR. The above equation means that, if it has a solution, then the Euclidean norm of the vector $\overline{a}$ must be in the same order as that of the vector $\overline{\xi}_{H}$. It can therefore be declared from Lemma \ref{lemma:1} that this equation is equivalent to that
\begin{equation}
\Gamma_{l}^{\perp}(\omega_{i}|_{i=1}^{N},\theta^{[0]})
\Omega(\omega_{i}|_{i=1}^{N},\theta^{[0]}) \overline{\xi}_{H}
=  O\left( ||\overline{\xi_{H}}||_{2}^{2} \right)
\label{eqn:data-err-11}
\end{equation}
which can be further expressed equivalently as
\begin{equation}
\overline{\xi}_{H} = S_{\!\! H} \widetilde{\xi}_{H} +   O\left( ||\widetilde{\xi}_{H}||_{2}^{2} \right)
\label{eqn:data-err-12}
\end{equation}
in which $\widetilde{\xi}_{H}$ is an arbitrary real vector with a compatible dimension.

On the other hand, when the matrix $\Gamma(\omega_{i}|_{i=1}^{N},\theta^{[0]})$ is of FCR, Eqs.(\ref{eqn:data-err-10}) and (\ref{eqn:data-err-12}) means that
\begin{equation}
\overline{a} = S_{\!\! A}(\omega_{i}|_{i=1}^{N},\theta^{[0]}) \widetilde{\xi}_{H} +   O\left( ||\widetilde{\xi}_{H}||_{2}^{2} \right)
\label{eqn:data-err-13}
\end{equation}

The proof can now be completed by putting Eqs.(\ref{eqn:data-err}), (\ref{eqn:data-err-8-b}), (\ref{eqn:data-err-12}) and (\ref{eqn:data-err-13}) together, recalling that
$\overline{\xi}_{H} = {\rm\bf col}\left\{\overline{\xi}_{Hi}|_{i=1}^{N} \right\}$, as well as $\overline{\Delta}_{i} = \overline{\Delta}_{i,r} + j \overline{\Delta}_{i,j}$ and
$\overline{\xi}_{Hi} = {\rm\bf vec}\left({\rm\bf col}\left\{\overline{\Delta}_{i,r},\; \overline{\Delta}_{i,j}\right\}\right)$ for each $i=1,2,\cdots,N$.   \hspace{\fill}$\Diamond$

\noindent\textbf{Proof of Theorem \ref{theo:6}:} For brevity, the dependence on the frequencies $\omega_{i}|_{i=1}^{N}$ and/or the parameter vector value $\theta^{[0]}$ is omitted for all the matrices and sets in this proof.

From the assumption that the set $\rm\bf\Theta$ is open, it can be declared that for each $k=1,2,\cdots,N$, there exists a positive number $\varepsilon_{k}^{[0]}$, such that for every positive number $\varepsilon$ satisfying  $\varepsilon \leq \varepsilon_{k}^{[0]}$, we have that $\theta^{[0]} +
S_{\!\! k}\:\xi \in {\rm\bf\Theta}$ for every $\xi \in \widehat{\rm\bf\Theta}_{F}(\varepsilon)$. On the other hand, when $\theta^{[0]} +
S_{\!\! k}\:\xi \in {\rm\bf\Theta}$ for each $\xi \in \widehat{\rm\bf\Theta}_{F}(\varepsilon)$, we have that the set  $\overline{\rm\bf\Theta}_{F}(\varepsilon,\omega_{i}|_{i=1}^{N},\theta^{[0]})$
is approximately to be a subset of the set ${\rm\bf\Theta}$, provided that $\varepsilon$ is sufficiently small.

Note that for each $\xi \in {\cal R}^{n_{s}}$, we have
\begin{displaymath}
\xi^{T}\left( \sum_{i=1}^{N} \widetilde{S}_{\!\! H,k}^{H} \widetilde{S}_{\!\! H,k} \right) \xi
=
\xi^{T}\left( \sum_{i=1}^{N} \left[\widetilde{S}_{\!\! H,k,r}^{T} \widetilde{S}_{\!\! H,k,r} + \widetilde{S}_{\!\! H,k,j}^{T} \widetilde{S}_{\!\! H,k,j} \right]\right) \xi
\end{displaymath}
It can therefore be declared that the problem of maximizing $\left|\left|S_{\!\! k}\:\xi\right|\right|_{2}^{2}$ under the constraint $\xi \in \widehat{\rm\bf\Theta}_{F}(\varepsilon)$ can be equivalently written as
\begin{equation}
\begin{array}{l}
\max \xi^{T} S_{\!\! k}^{T}S_{\!\! k}\:\xi \\
\hspace*{0.0cm}{\rm s.\:t.}\;
\xi^{T}\left( \sum_{i=1}^{N} \left[\widetilde{S}_{\!\! H,k,r}^{T} \widetilde{S}_{\!\! H,k,r} + \widetilde{S}_{\!\! H,k,j}^{T} \widetilde{S}_{\!\! H,k,j} \right]\right) \xi = \varepsilon
\end{array}
\label{a.17}
\end{equation}

Let $\mu$ be a Lagrange multiplier, and construct a Lagrangian function $J(\mu,\xi)$ as follows,
\begin{displaymath}
J(\mu,\xi) =  \frac{1}{2}\xi^{T} S_{\!\! k}^{T}S_{\!\! k}\:\xi - \frac{\mu}{2} \left\{
\xi^{T}\left( \sum_{i=1}^{N} \left[\widetilde{S}_{\!\! H,k,r}^{T} \widetilde{S}_{\!\! H,k,r} + \widetilde{S}_{\!\! H,k,j}^{T} \widetilde{S}_{\!\! H,k,j} \right]\right) \xi - \varepsilon \right\}
\end{displaymath}
Then the first order optimization condition with respect to the vector $\xi$ can be expressed as follows,
\begin{equation}
\frac{\partial J(\mu,\xi)}{\partial \xi} =  \left\{ S_{\!\! k}^{T}S_{\!\! k} - \mu \sum_{i=1}^{N} \left[\widetilde{S}_{\!\! H,k,r}^{T} \widetilde{S}_{\!\! H,k,r} + \widetilde{S}_{\!\! H,k,j}^{T} \widetilde{S}_{\!\! H,k,j} \right]\right\} \xi = 0
\label{a.18}
\end{equation}
meaning that the optimal $\xi$, denote by $\xi_{opt}^{[max]}$, is associated with a generalized eigenvalue of the matrix pair $\displaystyle \left(\sum_{i=1}^{N} \left[\widetilde{S}_{\!\! H,k,r}^{T} \widetilde{S}_{\!\! H,k,r} + \widetilde{S}_{\!\! H,k,j}^{T} \widetilde{S}_{\!\! H,k,j} \right],\;S_{\!\! k}^{T}S_{\!\! k} \right)$, which is given by Eq.(\ref{eqn:theo6}). In particular, let $\xi^{[i]}$ be an eigenvector associated with the $i$-th generalized eigenvalue $\mu^{[i]}$ of this matrix pair, $i=1,2,\cdots,n_{s}$. Then there exists a positive scalar $\alpha^{[i]}$, such that
\begin{equation}
\xi_{opt}^{[max]} = \alpha^{[i]} \xi^{[i]}
\end{equation}

Substitute this expression into the constraint of the optimization problem of Eq.(\ref{a.17}), it is obvious that the constraint is satisfied if and only if
\begin{equation}
\alpha^{[i]2} \xi^{[i]T}\left\{\sum_{i=1}^{N} \left[\widetilde{S}_{\!\! H,k,r}^{T} \widetilde{S}_{\!\! H,k,r} + \widetilde{S}_{\!\! H,k,j}^{T} \widetilde{S}_{\!\! H,k,j} \right]\right\} \xi^{[i]} = \varepsilon
\end{equation}
Hence
\begin{equation}
\alpha^{[i]} =\sqrt{ \frac{\varepsilon}
{\displaystyle \xi^{[i]T}\left\{\sum_{i=1}^{N} \left[\widetilde{S}_{\!\! H,k,r}^{T} \widetilde{S}_{\!\! H,k,r} + \widetilde{S}_{\!\! H,k,j}^{T} \widetilde{S}_{\!\! H,k,j} \right]\right\} \xi^{[i]}}}
\end{equation}
and
\begin{equation}
\xi_{opt}^{[max]} =\sqrt{ \frac{\varepsilon}
{\displaystyle \xi^{[i]T}\left\{\sum_{i=1}^{N} \left[\widetilde{S}_{\!\! H,k,r}^{T} \widetilde{S}_{\!\! H,k,r} + \widetilde{S}_{\!\! H,k,j}^{T} \widetilde{S}_{\!\! H,k,j} \right]\right\} \xi^{[i]}}} \xi^{[i]}
\label{eqn:sm-a}
\end{equation}
meaning that the Euclidean norm of the optimal vector, that is, $\xi_{opt}^{[max]}$, is in the same order of $\sqrt{\varepsilon}$.

Combining this equation with Eq.(\ref{a.18}), we further have that
\begin{eqnarray}
\xi_{opt}^{[max]T} {S}_{\!\! k}^{T} {S}_{\!\! k} \xi_{opt}^{[max]T}
& = & \!\!\alpha^{[i]2} \xi^{[i]T} {S}_{\!\! k}^{T} {S}_{\!\! k} \xi^{[i]T} \nonumber \\
&  & \hspace*{-2.2cm}=  \mu^{[i]}
\alpha^{[i]2} \xi^{[i]T}\left\{\sum_{i=1}^{N} \left[\widetilde{S}_{\!\! H,k,r}^{T} \widetilde{S}_{\!\! H,k,r} + \widetilde{S}_{\!\! H,k,j}^{T} \widetilde{S}_{\!\! H,k,j} \right]\right\} \xi^{[i]} \nonumber \\
&  & \hspace*{-2.2cm}= \mu^{[i]} \varepsilon
\end{eqnarray}
implying that the optimal value of the optimization problem of Eq.(\ref{a.17}) is equal to $\mu^{[1]} \varepsilon$, and the corresponding parameter vector value that reaches the maximum of $||\theta -\theta^{[0]}||_{2}$, denote it by $\xi_{opt}^{[1]}$, is given by Eq.(\ref{eqn:sm-a}) with $\xi^{[i]}$ being fixed to be $\xi^{[1]}$.

For each $k=2,3,\cdots,n_{s}$, formulate recursively the following optimization problem,
\begin{displaymath}
\begin{array}{l}
\max \xi^{T} S_{\!\! k}^{T}S_{\!\! k}\:\xi \\
\hspace*{0.0cm}{\rm s.\:t.}\;
\xi^{T}\left( \sum_{i=1}^{N} \left[\widetilde{S}_{\!\! H,k,r}^{T} \widetilde{S}_{\!\! H,k,r} + \widetilde{S}_{\!\! H,k,j}^{T} \widetilde{S}_{\!\! H,k,j} \right]\right) \xi = \varepsilon \\
\hspace*{0.50cm} \xi \in {\cal R}^{n_{s}}\backslash {\rm\bf span}\left\{\xi_{opt}^{[1]},\;\xi_{opt}^{[2]},\; \cdots,\; \xi_{opt}^{[k-1]} \right\}\end{array}
\end{displaymath}
Then on the basis of the properties of singular values and the above arguments, it can also be shown that the optimal value of this  optimization problem is equal to $\mu^{[k]} \varepsilon$, while the associated parameter vector value that reaches the maximum of $||\theta -\theta^{[0]}||_{2}$, denote it by $\xi_{opt}^{[k]}$, is given by Eq.(\ref{eqn:sm-a}) with $\xi^{[i]}$ being fixed to be $\xi^{[k]}$.

The proof can now be completed by the definition of the set $\overline{\rm\bf\Theta}_{F}(\varepsilon,\omega_{i}|_{i=1}^{N},\theta^{[0]})$, as well as those of the absolute and relative sloppiness metrics
${\rm\bf Sm}^{[a]}_{F}(\varepsilon,\omega_{i}|_{i=1}^{N},\theta^{[0]})$ and ${\rm\bf Sm}^{[r,k]}_{F}(\varepsilon,\omega_{i}|_{i=1}^{N},\theta^{[0]})$ of the descriptor system $\rm\bf\Sigma$.   \hspace{\fill}$\Diamond$



\begin{small}

\end{small}

\end{document}